\documentclass[aps,prd,nofootinbib,floatfix,superscriptaddress,secnumarabic]{revtex4}
\usepackage{hyperref}
\usepackage{amsmath}
\usepackage{bm}
\usepackage{graphicx}
\usepackage{epsf}
\usepackage{epsfig}
\usepackage{amsmath}
\usepackage{amsfonts,amssymb}
\usepackage{colordvi}
\usepackage{color}
\usepackage{lscape}
\usepackage{rotfloat}
\usepackage{rotating}

\usepackage{dsfont}
\usepackage{slashed}
\usepackage{booktabs}

\newcommand{\be}{\begin{equation}}
\newcommand{\ee}{\end{equation}}
\newcommand{\bea}{\begin{eqnarray}}
\newcommand{\eea}{\end{eqnarray}}

\newcommand{\MSbar}{{\overline{\rm MS}}}

\newcommand{\Dslash}{{\not{\hspace{-0.1cm}D}}}

\newcommand{\pa}{\partial}
\newcommand{\la}{\lambda}
\def\slashed{{/}\mskip-10.0mu}

\begin{document}

\title{Renormalization and Mixing of the Gluino-Glue Operator on the Lattice}
\author{M.~Costa\footnote[1]{ kosta.marios@ucy.ac.cy, herodotou.herodotos@ucy.ac.cy,  philippides.phivos@ucy.ac.cy, panagopoulos.haris@ucy.ac.cy }}
\affiliation{Department of Physics, University of Cyprus, Nicosia, CY-1678, Cyprus}
\affiliation{Department of Mechanical Engineering and Materials Science and Engineering, Cyprus University of Technology, Limassol, CY-3036, Cyprus}
\author{H.~Herodotou}
\affiliation{Department of Physics, University of Cyprus, Nicosia, CY-1678, Cyprus}
\author{P. Philippides} 
\affiliation{Department of Physics, University of Cyprus, Nicosia, CY-1678, Cyprus}
\author{H.~Panagopoulos}
\affiliation{Department of Physics, University of Cyprus, Nicosia, CY-1678, Cyprus}

\begin{abstract}
We study the mixing of the Gluino-Glue operator in ${\cal N}$=1 Supersymmetric Yang-Mills theory (SYM), both in dimensional regularization and on the lattice. We calculate its renormalization, which is not only multiplicative, due to the fact that this operator can mix with non-gauge invariant operators of equal or, on the lattice, lower dimension. These operators carry the same quantum numbers under Lorentz transformations and global gauge transformations, and they have the same ghost number.

We compute the one-loop quantum correction for the relevant two-point and three-point Green's functions of the Gluino-Glue operator. This allows us to determine renormalization factors of the operator in the $\MSbar$ scheme, as well as the mixing coefficients for the other operators. To this end our computations are performed using dimensional and lattice regularizations. We employ a standard discretization where gluinos are defined on lattice sites and gluons reside on the links of the lattice; the discretization is based on Wilson's formulation of non-supersymmetric gauge theories with clover improvement. The number of colors, $N_c$, the gauge parameter, $\beta$, and the clover coefficient, $c_{\rm SW}$, are left as free parameters.
\end{abstract}
\maketitle

\section{Introduction}
Supersymmetry (SUSY) has a long history as a viable extension of the Standard Model~\cite{Martin:1997ns, Quevedo:2010ui, Zyla:2020}. It provides possible answers to a number of open questions, such as the hierarchy problem, a candidate for dark matter, and a scenario for grand unification; its presence is also compelling in the context of String Theory. Experimental signatures of Supersymmetry have thus far been elusive, despite decades of search in large-scale experiments, including recent findings at LHC. Nevertheless, there is a major ongoing research effort in this direction, see e.g.~\cite{Santra:2020mfi, CMS:2019tlp}, given that no satisfactory solution to the above open questions has come about to date. In order for SUSY to be compatible with ``low-energy'' phenomenology, it is expected that it must be spontaneously broken in nature. A detailed study of spontaneous breaking must necessarily rely on nonperturbative methods, thus calling for an investigation within lattice field theory~\cite{Curci:1986sm, Creutz:2001, Giet&Poppitz,Kaplan:2009, Catterall:2014vga, Joseph:2015xwa, Ali:2018fbq, Endrodi:2018ikq, Giedt:2009yd, Bergner:2016sbv,Ali:2020mvj}. To date the study of supersymmetric models on the lattice has been very limited, due to their sheer complexity. The fact that SUSY is broken explicitly on the lattice poses severe issues to its correct simulation and to the numerical study of spontaneous SUSY breaking. A thorough renormalization procedure is an essential prerequisite towards non-perturbative investigations. This procedure must determine all relevant renormalization and mixing coefficients in the Lagrangian, so that the correct continuum limit can be reached, with SUSY and chiral symmetry restored in this limit~\cite{Giet&Poppitz, Giedt:2009yd}. 

A most appropriate prototype theory, exhibiting all the above features and including both gauge and matter fields, is Supersymmetric Quantum Chromodynamics (SQCD). The study of SQCD is already very complicated on the lattice due to its many degrees of freedom and interaction terms~\cite{Costa:2017rht, Costa:2018mvb}. Consequently, the study of composite operators and their mixing is presently out of reach, especially at the nonperturbative level. A simpler theory, and an important forerunner to the more complex models, is the Supersymmetric Yang-Mills theory (SYM). It contains only gauge fields and it exhibits an interesting spectrum of bound states, in particular particles made of gluino ($\lambda$) and gluon $(u_\mu)$ fields. Preliminary nonperturbative investigations in this direction were performed in Refs.~\cite{Ali:2019agk,Ali:2018dnd,Ali:2020mvj}. A fundamental ingredient in these investigations is the ``Gluino-Glue'' composite operator, ${\cal O}_{Gg}$. In the present work we study thoroughly the renormalization and mixing of this operator, to one loop in perturbation theory. 

The Gluino-Glue operator is a composite operator made up of a gluon and a gluino field; it is thus flavor-singlet, and it has the lowest possible dimensionality (7/2) compatible with gauge invariance. It defined as\footnote{${\rm{tr}}_c$ means trace over color matrices.}:
\be
{\cal O}_{Gg} = \sigma_{\mu \nu} \,{\rm{tr}}_c (\, u_{\mu \nu} \lambda )
\label{GgO}
\ee
where:
\be
 \sigma_{\mu \nu}=\frac{1}{2} [\gamma_{\mu},\gamma_{\nu}],\quad  u_{\mu \nu} = \partial_{\mu}u_{\nu}-\partial_{\nu}u_{\mu}+ i g [u_{\mu},u_{\nu}].
\ee
Acting on the vacuum, ${\cal O}_{Gg}$ is expected to excite a light bound state of the theory, which is a potential supersymmetric partner of the glueballs and the gluinoballs~\cite{VEN}. 

Within the SYM formulation, we compute the relevant two-point and three-point Green's functions of the Gluino-Glue operator with external gluino, gluon and ghost fields, using both dimensional regularization and lattice regularization. Quantum corrections cause mixing of some non-gauge invariant operators which have the same quantum numbers as ${\cal O}_{Gg}$. As in non-supersymmetric theories, these operators are separated in three classes~\cite{Collins:1984xc, Collins:1994ee}. The Gluino-Glue operator belongs to a separate class by itself since there are no other gauge-invariant operators of equal and lower dimensionality which can mix with ${\cal O}_{Gg}$. The renormalization of ${\cal O}_{Gg}$ as well as the corresponding mixing coefficients are calculated in the $\MSbar$ scheme.

This paper is organized as follows. Section~\ref{sec2} shows all relevant definitions and all operators which could possibly mix with ${\cal O}_{Gg}$. Section~\ref{sec3} describes the calculation setup. In Section~\ref{sec4}, we present our results for the Green's functions, the renormalization factors as well as the mixing coefficients in dimensional regularization. Section~\ref{sec5} introduces the lattice action. We use clover fermions and Wilson gluons. We compute all relevant Green's functions of ${\cal O}_{Gg}$ within lattice perturbation theory. We also present the renormalization factors and mixing coefficients in the lattice regularization and the $\MSbar$ scheme. Finally, we conclude in Section~\ref{sec6} with a discussion of our results and possible future extensions of our work. For completeness, we have included an Appendix containing the one-loop renormalization factors for the gluon ($Z_u$) and gluino ($Z_\la$) fields. Results for the latter quantities, and for other renormalization factors that we need here, have been already presented in Ref.~\cite{Costa:2017rht} for different discretizations.

\section{Definitions and candidate operators of dimension 7/2 and 5/2}
\label{sec2}
In this Section we briefly introduce the notation used in this paper and we present all candidate operators that may mix with ${\cal O}_{Gg}$.
The action of SYM in Minkowski space is ($D^{\alpha}$ is an auxiliary field):
\begin{equation} \label{SYMD}
\mathcal{L}_{SYM}=-\frac{1}{4}u_{\mu \nu}^{\alpha}u_{\mu \nu}^{\alpha}+\frac{i}{2}\bar{\lambda}^{\alpha}_{M}\gamma^{\mu}\mathcal{D}_{\mu}\lambda^{\alpha}_{M}+\frac{1}{2}D^{\alpha}D^{\alpha},  \quad\quad \la_M = \left( {\begin{array}{c} \la_a\\ \bar \la^{\dot a} \end{array} } \right)
\end{equation}
The subscript $M$ recalls the Majorana nature of the gluino. Henceforth we will omit this subscript for simplicity. The field strength $u_{\mu \nu}$ and the covariant derivarive of $\la$ are:
\bea
\mathcal{D}_{\mu}\lambda^{\alpha}&=&\partial_{\mu}\lambda^{\alpha}-g f^{\alpha \beta \gamma}u_{\mu}^{\beta}\lambda^{\gamma}\nonumber\\
u_{\mu \nu}^{\alpha}&=&\partial_{\mu}u_{\nu}^{\alpha}-\partial_{\nu}u_{\mu}^{\alpha}-g f^{\alpha \beta \gamma}u_{\mu}^{\beta}u_{\nu}^{\gamma}
\eea

By eliminating the auxiliary field, we get:
\begin{equation} \label{SYMEq1}
\mathcal{L}_{SYM}=-\frac{1}{4}u_{\mu \nu}^{\alpha}u_{\mu \nu}^{\alpha}+\frac{i}{2}\bar{\lambda}^{\alpha}\gamma^{\mu}\mathcal{D}_{\mu}\lambda^{\alpha}
\end{equation}
where the Lagrangian, ${\cal L}_{\rm SYM}$, is invariant up to a total derivative under the supersymmetry transformations with Grassmann parameter $\xi$:
\bea
\delta_\xi u_\mu^{\alpha} & = & -i \bar \xi \gamma^\mu \lambda^{\alpha}, \nonumber \\
\delta_\xi \lambda^{\alpha} & = & \frac{1}{4} u_{\mu \nu}^{\alpha} [\gamma^{\mu},\gamma^{\nu}] \xi \,.
\label{susytransfDirac}
\eea

Gauge trasformations act on the fields as:
\begin{align}
u'_\mu &= G^{-1} u_\mu G + \frac{i}{g} (\partial_\mu G^{-1})G, &\lambda' &= G^{-1} \lambda G &
\label{SgaugeTranComponents}
\end{align}
where $G(x) \equiv e^{i \omega^{\alpha}(x) T^{\alpha}}$, $T^\alpha$ are the generators of $su(N)$, and $\omega^\alpha(x)$ are real parameters.

Given that the renormalized theory does not depend on the choice of a gauge fixing term, and given that many regularizations, in particular the lattice regularization, violate supersymmetry at intermediate steps, one may as well choose the standard covariant gauge fixing term, proportional to $(\partial_\mu u^\mu)^2$, rather than a supersymmetric variant~\cite{Miller:1983pg, Costa:2017rht}. The full SYM action thus includes a gauge-fixing term and a ghost term arising from the Faddeev-Popov procedure:
\begin{equation}
{\cal S}_{GF}= \frac{1}{1 - \beta}\,\int d^4x \, \frac{1}{2} \left( B^\alpha \, B^\alpha \right),\quad B^\alpha \equiv \partial^\mu u_\mu^\alpha
\label{sgf}
\end{equation}
where $\beta$ is the gauge parameter ($\beta=1(0)$ corresponds to Landau (Feynman) gauge), and 
\begin{equation}
{\cal S}_{Ghost}= - 2 \int d^4x \,{\rm{tr}}_c \left( \bar{c}\, \partial^{\mu}D_\mu  c\right).
\label{sghost}
\end{equation}
The ghost field $c$ is a Grassmann scalar which transforms in the adjoint representation of the gauge group, and: ${\cal{D}}_\mu c =  \pa_{\mu} c + i g \,[u_\mu,c]$. Consequently, the the total action in the continuum has the form:
\begin{equation}
{\cal S}_{\rm total SYM} = {\cal S}_{\rm SYM} + {\cal S}_{GF} + {\cal S}_{Ghost}.
\label{ScontALL}
\end{equation}
By construction, ${\cal S}_{\rm total SYM}$ is not gauge invariant; however it is invariant under Becchi-Rouet-Stora-Tyutin (BRST) transformations. The latter involve parameters that take their values in a Grassmann algebra. The BRST trasformations for the fields of the full SYM action can be found by setting $\omega^{\alpha}$ in Eq.~(\ref{SgaugeTranComponents}) equal to $c^{\alpha} \xi$, where $\xi$  is a Grassmann variable. Thus, the fields appearing in Eq.~(\ref{ScontALL}) behave as follows:
\bea
u^{\alpha}_{\mu}&\rightarrow& u^{\alpha}_{\mu}+(\partial_{\mu}c^a+g f^{\alpha \beta \gamma}c^{\beta}u^{\gamma}_{\mu})\ \xi, \nonumber \\
\lambda &\rightarrow & \lambda+ gc^{\alpha}\lambda^{\beta}f^{\alpha\beta\gamma}T^{\gamma}\ \xi\nonumber \\
c^\alpha &\rightarrow & c^\alpha -\frac{g}{2}f^{\alpha\beta\gamma}c^\beta c^\gamma \ \xi, \nonumber \\
\bar{c}^\alpha &\rightarrow & \bar{c}^\alpha +B^\alpha \ \xi,\nonumber \\
B^\alpha &\rightarrow & B^\alpha,
\label{ffb}
\eea
Under these transformations, the action is indeed invariant. Given that the effect of a BRST transformation on fields is that of a gauge transformation, all gauge invariant parts of the action will automatically also be BRST invariant. 

By general renormalization theorems, the operators that will possibly mix with ${\cal O}_{Gg}$ are either gauge invariant (class G) or belong to one of three classes. Class A operators are the BRST variation of other operators. Class B operators vanish by the equations of motion. Lastly, class C contains all other operators with compatible quantum numbers. 

In SYM, there are no further gauge invariant operators with the same quantum numbers as ${\cal O}_{Gg}$. Let us now determine the members of class A, B and C. By Eq.~(\ref{ffb}), the operators whose BRST variation will be the members of class A must necessarily have the same index structure as ${\cal O}_{Gg}$, i.e., one free spinor index and no free color or Lorentz indices; in addition, their dimensionality must not exceed $5/2$. This requirement leaves only two candidates:
\be
\delta_{BRST}\left(\la^{\alpha} \bar c^{\alpha}\right)= \la^{\alpha}B^{\alpha}\xi + gf^{\alpha \beta \gamma}c^{\beta}\la^{\gamma}\bar{c}^{\alpha}\xi \Rightarrow {\cal O}_{A1}\equiv \la^{\alpha}B^{\alpha} + gf^{\alpha \beta \gamma}c^{\beta}\la^{\gamma}\bar{c}^{\alpha}
\label{BRST1op}
\ee
\be
\delta_{BRST}\left(\la^{\alpha} c^{\alpha}\right) = f^{\alpha \beta \gamma}c^{\alpha}c^{\beta}\la^{\gamma}\xi
\label{BRST2op}
\ee
We note that BRST variations of operators are automatically BRST invariant. Operators containing unequal numbers of ghost and antighost fields cannot mix with ${\cal O}_{Gg}$, since ${\cal O}_{Gg}$ has ghost number zero. Thus, the only admissible Class A operator is ${\cal O}_{A1}$, which is written in Eq.~(\ref{BRST1op}). Class A operators have vanishing matrix elements in physical external states with transverse polarization. However, they must be correctly taken into account for the renormalization of ${\cal O}_{Gg}$. Similar comments apply to class B and C. The second term of ${\cal O}_{A1}$, will appear also in class C (see below): ${\cal O}_{C4} = g f^{\alpha \beta \gamma}c^{\beta}\la^{\gamma}\bar{c}^{\alpha}$. In order to find the mixing coefficient for ${\cal O}_{C4}$, we will have to calculate the three-point Green's function shown in the diagrams of Fig.~\ref{fig3ptgCC}.

For the class B operators we check the equations of motion for the gluino and gluon fields. Taking into account that operators must have zero ghost number and that the gluon equation of motion has already dimension 3, we conclude that only the gluino equation of motion may contribute; we must also multiply it by a factor of $u_\mu \gamma_\mu$ in order to render it colorless. This leads to only one member in class B: ${\cal O}_{B1} = {\rm{tr}}_c (\slashed{u} \Dslash \la )$.

Class C operators are neither gauge invariant, nor BRST variations, nor operators that vanish by the equations of motion; but they have the correct free indices, dimensionality and ghost number.

We present all candidate operators which can mix with ${\cal O}_{Gg}$:
\bea
{\cal O}_{A1} &=& {\rm{tr}}_c (\la B ) - ig\, {\rm{tr}}_c(\la [c, \bar{c}])\\[2ex]
{\cal O}_{B1} &=&  {\rm{tr}}_c (\slashed{u} \Dslash \la )\\[2ex]
{\cal O}_{C1} &=& {\rm{tr}}_c (\partial_{\mu}\la\, u^{\mu}) \\[2ex]
{\cal O}_{C2} &=& {\rm{tr}}_c (\slashed{u} \la) \\[2ex]
{\cal O}_{C3} &=& ig\,\sigma_{\mu \nu} {\rm{tr}}_c (\, \lambda [u_{\mu}, u_{\nu}] )\\[2ex]
{\cal O}_{C4} &=& ig\, {\rm{tr}}_c(\la [c, \bar{c}])
\label{All_operators}
\eea

In the context of SQCD~\cite{Costa:2018mvb}, there is a plethora of further operators which mix; they all share the same quantum numbers, including being flavor singlet and having baryon number zero, containing also quark and squark fields.

Class C operators cannot contribute in the continuum for the purpose of $\MSbar$-renormalization. However, they may give finite mixing coefficients on the lattice. Note also that the operator ${\cal O}_{C2}$ is of lower dimension and it will not mix with ${\cal O}_{Gg}$ in dimensional regularization; it may however show up in the lattice formulation. The presence of symmetries, which are preserved by the SYM action, both in the continuum and on the lattice, forbids other operators from mixing with the Gluino-Glue operator.

\section{Calculation setup}
\label{sec3}

The renormalization coefficients of all candidate operators are calculated by constructing a $7 \times 7$ mixing matrix, which includes: a gauge invariant operator, ${\cal O}_{Gg}$, a BRST invariant operator, ${\cal O}_{A1}$, an operator that vanishes by the equations of motion, ${\cal O}_{B1}$, and four class C operators: ${\cal O}_{C1}$, ${\cal O}_{C2}$, ${\cal O}_{C3}$, ${\cal O}_{C4}$. The mixing matrix relates the renormalized operators to the bare ones. It was checked that the divergent parts of the mixing matrix have a block-triangular form\footnote{i.e. an operator from class G, A, B, C can mix with operators from the same or from the subsequent classes but not from the previous classes.}. We calculate only its first row since we are interested in the renormalization of the Gluino-Glue operator. Thus, the renormalized operator ${\cal O}^R_{Gg}$ is related to the bare ones, through:	
\begin{equation}
{\cal O}^R_{Gg}  = Z_{Gg} {\cal O}^B_{Gg} + z_{A1}{\cal O}^B_{A1} + z_{B1} {\cal O}^B_{B1}  + z_{C1}{\cal O}^B_{C1} + z_{C2}{\cal O}^B_{C2} + z_{C3}{\cal O}^B_{C3} + z_{C4}{\cal O}^B_{C4}
\label{Zz} 
\end{equation}
where the renormalization factor $Z = \openone + {\cal O}(g^2)\,$ and the mixing coefficients $z = {\cal O}(g^2)$ should more properly be denoted as $Z^{X,Y}$and $z^{X,Y}$, where $X$ is the regularization and $Y$ the renormalization scheme. Superscript $B$ stands for bare and $R$ for renormalized quantities. 

As an example, if one is interested in the full mixing matrix in $DR$ and the $\MSbar$ renormalization scheme, its explicit form is triangular. We have omitted class C operators since their mixing coefficients, if they appear, will be finite and thus they will not contribute in the $\MSbar$ scheme:
\begin{gather}
 \begin{pmatrix} {\cal O}^R_{Gg} & \\[2ex] {\cal O}^R_{A1}\\[2ex] {\cal O}^R_{B1} \end{pmatrix}
 =
\begin{pmatrix}
   Z_{Gg} &
   z_{A1} &
   z_{B1}  \\[2ex]
   0  &
   Z_{A1} &
   z_{A B}  \\[2ex]
   0  &
   0 &
   Z_{B1}  \\[2ex]
   \end{pmatrix}
  \begin{pmatrix} {\cal O}^B_{Gg} & \\[2ex] {\cal O}^B_{A1} \\[2ex] {\cal O}^B_{B1} \end{pmatrix}
\end{gather}
From the above matrix, it is clear that class B operators can only mix with operators of the same class (in $DR$ and $\MSbar$). If the renormalization matrix were not triangular then the renormalized operators of Classes A and B would not vanish on-shell, even though the bare operators vanish. Triangularity ensures that for matrix elements in physical states $|P\rangle$, $|P'\rangle$, we have:
\be
\langle  P|{\cal O}^R_{Gg}|P'\rangle  = Z_{Gg} \langle  P|{\cal O}^B_{Gg}|P'\rangle
\ee
Thus, one can ignore non-gauge invariant operators for physical states. On the other hand, if one calculates a Green's function with elementary external fields (as is typically done for deducing nonperturbative renormalization on the lattice), they may have finite contributions which cannot be ignored even in the $\MSbar$ scheme.

In order to calculate the one-loop renormalization factor and the mixing coefficients, we compute the two-point Green's function of ${\cal O}^R_{Gg}$ with one external gluino and one external gluon fields, 
as well as three-point Green's functions with external gluino/gluon/gluon fields 
and with external gluino/ghost/antighost fields. Furthermore, renormalization conditions involve the renormalization factors of the gluon, gluino, ghost and coupling constant. For completeness, we present the definitions of these factors:

\bea
u_{\mu}^R &=& \sqrt{Z_u}\,u^B_{\mu},\\
\la^R &=& \sqrt{Z_\la}\,\la^B,\\
c^R &=& \sqrt{Z_c}\,c^B, \\
g^R &=& Z_g\,\mu^{-\epsilon}\,g^B, 
\eea
where $\mu$ is an arbitrary scale with dimensions of inverse length. For one-loop calculations, the distinction between $g^R$ and $\mu^{-\epsilon}\,g^B$ is inessential in many cases; we will simply use $g$ in those cases. Our results are presented as functions of the $\MSbar$ scale $\bar\mu$ which is related to $\mu$ through\footnote{$\gamma_E$  is Euler's constant: $\gamma_E = 0.57721\ldots$\, .}: $\mu = \bar \mu \sqrt{e^{\gamma_E}/ 4\pi}$.

\begin{figure}[ht!]
\centering
\includegraphics{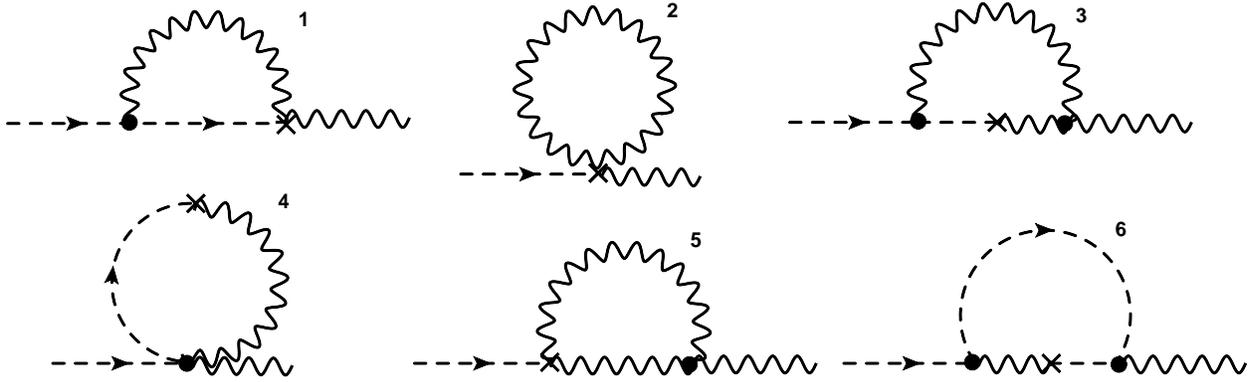}
\caption{One-loop Feynman diagrams contributing to the two point Green's
  function of the Gluino-Glue operator, $\langle u_\nu {\cal O}_{Gg} \bar \la  \rangle$\,.  A wavy (dashed) line represents gluons (gluinos). A cross denotes the insertion of the Gluino-Glue operator. Diagrams 2, 4 do not appear in dimensional regularization; they do however show up in the lattice formulation.}
\label{fig2pt}
\end{figure}

All of our results are computed as functions of the coupling constant $g$, the number of colors $N_c$, the gauge fixing parameter $\beta$, the clover parameter $c_{\rm SW}$ and the external momenta $q_i$. More specifically, we calculate the two-point Green's function $\langle u_\nu^{\alpha_1}(-q_1) {\cal O}_{Gg}(x) \bar \la^{\alpha_2}(q_2)  \rangle$, for three choices of the external momenta $q_1$ and $q_2$. This has been done in order to differentiate among the tree-level structures of the operators containing a gluon and a gluino field. Clearly, all operators that can possibly mix with ${\cal O}_{Gg}$ appear on the rhs of Eq.~(\ref{Zz}); the tree-level Green's functions of these operators naturally show up in the results for the one-loop Green's functions of ${\cal O}^R_{Gg}$, thus allowing us to deduce the corresponding mixing coefficients. The one-loop Feynman diagrams (one-particle irreducible (1PI)) contributing to this Green's function are shown in Fig.~\ref{fig2pt}. 

\begin{figure}[ht!]
\centering
\includegraphics{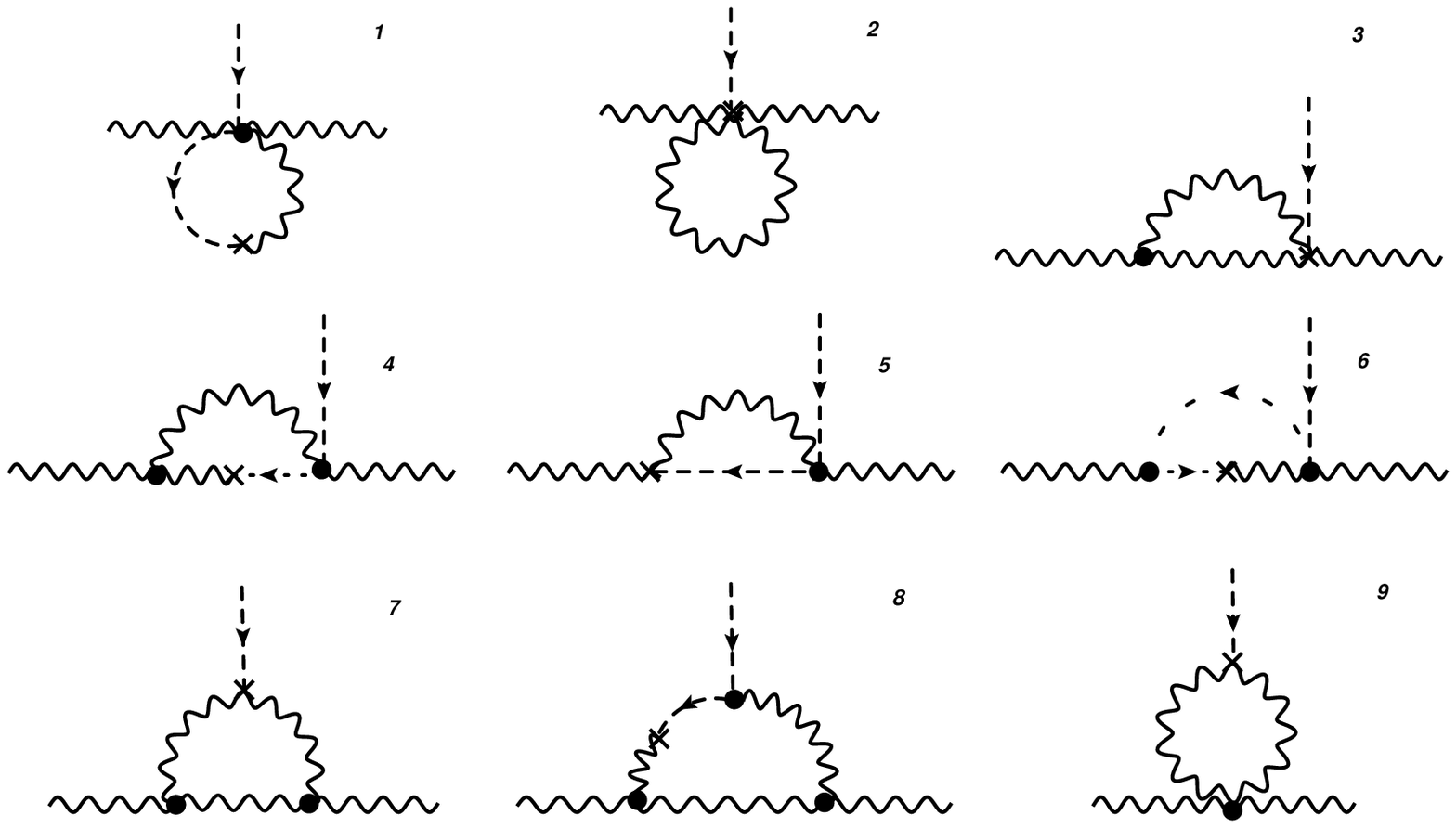}\\
\includegraphics{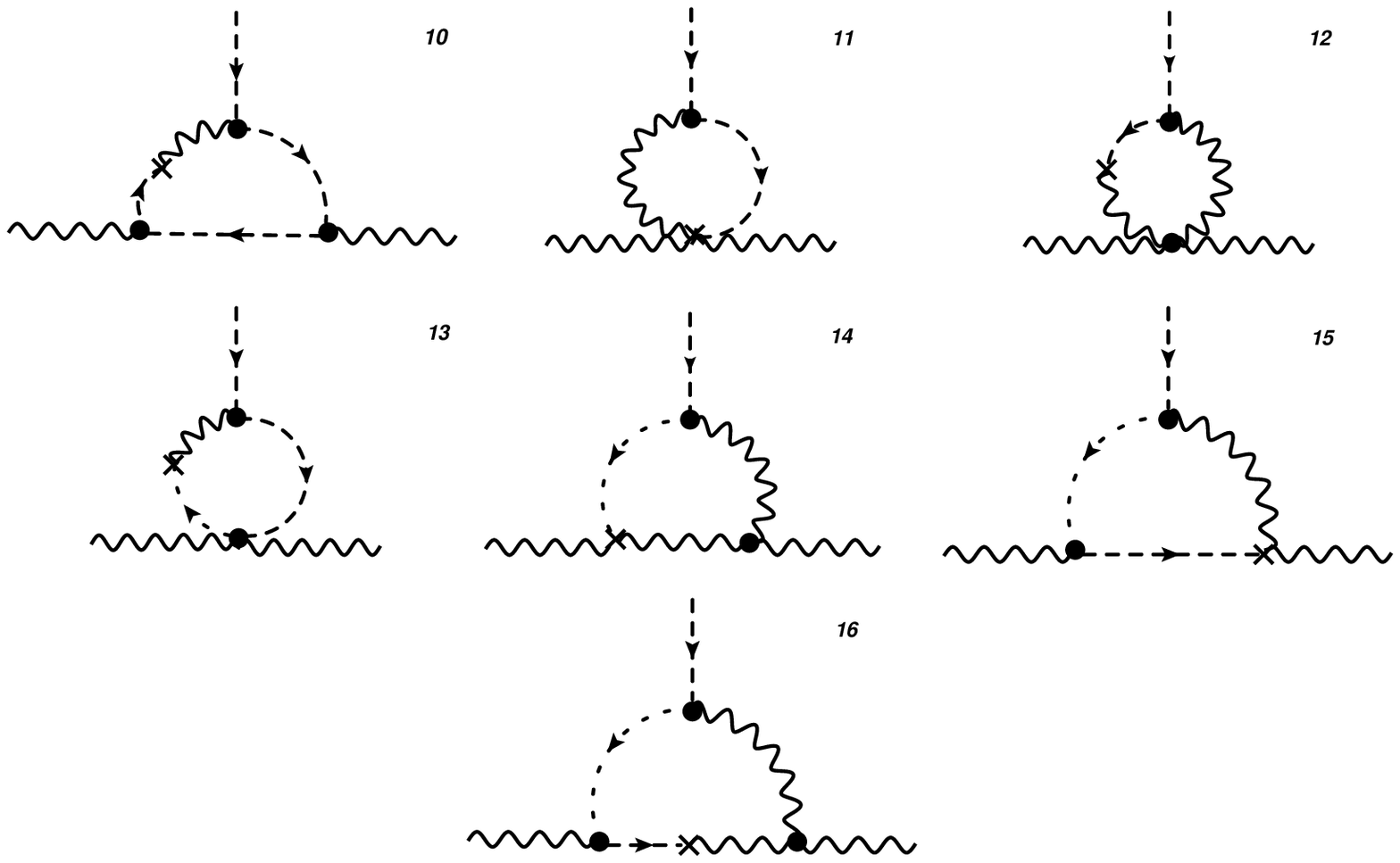}
\caption{One-loop Feynman diagrams contributing to the three point Green's
  function of the Gluino-Glue operator, $\langle u_\nu u_\mu {\cal O}_{Gg}\bar \la  \rangle$\,.  A wavy (dashed) line represents gluons (gluinos). Diagrams 1, 2, 3, 5, 6, 11, and 13 do not appear in dimensional regularization but they contribute in the lattice regularization. A cross denotes the insertion of the operator. A mirror version (under exchange of the two external gluons) of diagrams 3, 4, 5, 6, 8, 10, 14, 15 and 16 must also be included.}
\label{fig3ptguu}
\end{figure}

\begin{figure}[ht!]
\centering
\includegraphics{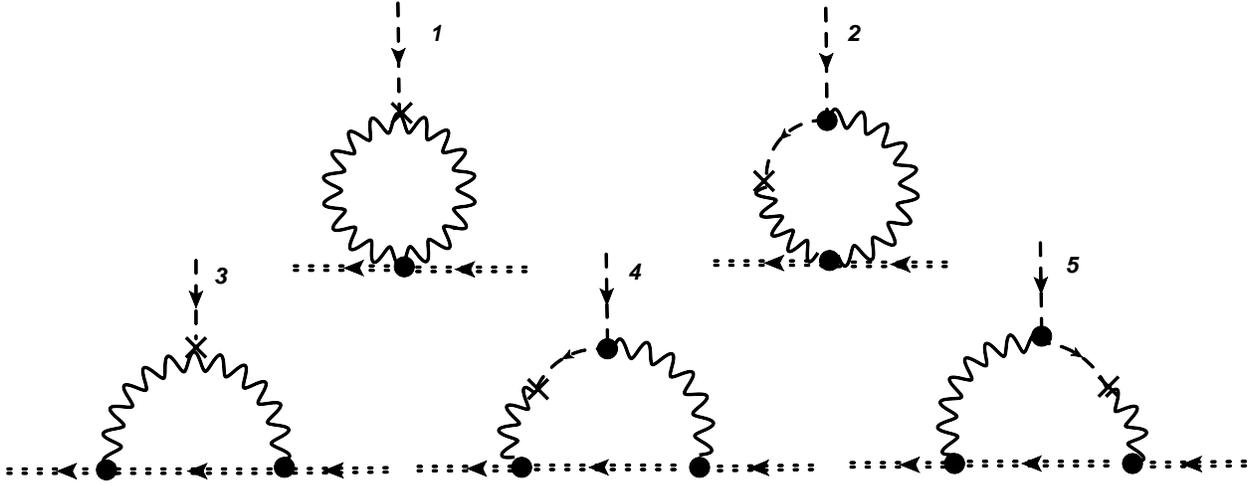}
\caption{One-loop Feynman diagrams contributing to the three point Green's
  function of the Gluino-glue operator, $\langle c \, {\cal O}_{Gg} \, \bar c \, \bar \la \rangle$\,.  A wavy (dashed) line represents gluons (gluinos). A cross denotes the insertion of the operator. The ``double dashed'' line is the ghost field. Diagrams 1 and 2 do not appear in dimensional regularization; they do however show up in the lattice formulation.}
\label{fig3ptgCC}
\end{figure}

We also calculate the three-point functions $\langle u_\nu^{\alpha_1}(-q_1) \,u_\mu^{\alpha_2}(-q_2) \,{\cal O}_{Gg}(x)\, \bar\lambda^{\alpha_3}(q_3) \rangle$ and $\langle c^{\alpha_3}(q_3) \,{\cal O}_{Gg}(x)\,\bar c^{\alpha_2}(q_2) \bar\lambda^{\alpha_1}(q_1) \rangle$, corresponding to the Feynman diagrams shown in Fig.~\ref{fig3ptguu} and Fig.~\ref{fig3ptgCC} in order to determine the mixing coefficients with ${\cal O}_{C3}$ and ${\cal O}_{C4}$, respectively. We present below the results of each three-point function in a given choice of the external momenta $q_1$, $q_2$ and $q_3$. Even though mixing is not expected to appear in the case of $DR$, we use this fact as a check on our perturbative results in the continuum. In the lattice regularization we expect finite mixing with these operators. In fact we have seen that there is no mixing with ${\cal O}_{C4}$ but on the lattice finite mixing with ${\cal O}_{C3}$ emerges.

Since renormalization conditions are typically imposed on amputated renormalized Green's functions, let us relate the latter to the bare ones. 

For the gluino-gluon Green's function:
\bea 
\langle u_\nu^R \,{\cal O}^R_{Gg}\, \bar\lambda^R \rangle _{amp} &=& Z_\la^{-1/2} \,Z_u^{-1/2} Z_{Gg} \langle u_\nu^B \,{\cal O}^B_{Gg}\,\bar\lambda^B \rangle _{amp}\nonumber\\
&+& z_{A1}  \langle u_\nu^B \, {\cal O}^B_{A1}\, \bar\lambda^B \rangle _{amp}^{tree} + z_{B1} \langle  u_\nu^B \, {\cal O}^B_{B1} \, \bar\lambda^B \rangle _{amp}^{tree}\nonumber\\ 
&+& z_{C1} \langle u_\nu^B \,{\cal O}^B_{C1} \, \bar\lambda^B \rangle _{amp}^{tree}+ z_{C2}\langle u_\nu^B  \, {\cal O}^B_{C2}\, \bar\lambda^B \rangle _{amp}^{tree} + {\cal O}(g^4)
\label{2ptGFexpr}
\eea

Similarly for the gluino-gluon-gluon Green's function:
\bea 
\langle u_\nu^R\,u_\mu^R \,{\cal O}^R_{Gg}\, \bar\lambda^R \rangle _{amp} &=& Z_\la^{-1/2} \,Z_u^{-1} Z_{Gg} \langle u_\nu^B u_\mu^B \,{\cal O}^B_{Gg}\, \bar\lambda^B  \rangle _{amp}\nonumber\\
&+&  z_{B1}\langle u_\nu^B u_\mu^B \, {\cal O}^B_{B1}\, \bar\lambda^B \rangle _{amp}^{tree} + z_{C3}\langle u_\nu^B u_\mu^B \, {\cal O}^B_{C3}\, \bar\lambda^B \rangle _{amp}^{tree} + {\cal O}(g^4)
\label{eptGFexprGGg}
\eea
We should renormalize the coupling constant in the tree-level three point Green's function of ${\cal O}^B_{Gg}$, thus we multiply it by $Z_g^{-1}$ since the relevant ${\cal O}^B_{Gg}$ vertex contains one power of $g^B$. Given that these calculations are up to one-loop order, the coupling constant in the one-loop bare Green's function (being higher order in $g$) is already expressed in terms of the renormalized coupling.

Lastly, for the gluino-ghost-antighost Green's function:
\bea 
\langle c^R \,{\cal O}^R_{Gg}\,\bar c^R\, \bar\lambda^R \rangle _{amp} &=&  
 Z_c^{-1} \,Z_\la^{-1/2} Z_{Gg} \langle c^B \,{\cal O}^B_{Gg}\,\bar c^b \bar\lambda^B  \rangle _{amp}\nonumber\\
&+&z_{A1}\langle c^B \, {\cal O}^B_{A1}\,\bar c^B \bar\lambda^B \rangle _{amp}^{tree} + z_{C4}\langle c^B \, {\cal O}^B_{C4}\,\bar c^B \bar\lambda^B \rangle _{amp}^{tree} + {\cal O}(g^4)
\label{3ptGFexpr2}
\eea

A few comments are in order here:
\begin{itemize}
\item[1] The gluino field and the gluon field renormalization factors, $Z_\la$ and $Z_u$ do not depend on flavour since this study is within the SYM theory. The ghost field renormalization constant, $Z_c$, is the same as in Ref.~\cite{Costa:2017rht}. In addition, continuum results for the renormalization factors of fields are also the same as in Ref.~\cite{Costa:2017rht}, setting $N_f=0$. The lattice results here have additional terms due to the fact that we use clover fermions.
\item[2] To avoid heavy notation we have omitted coordinate/momentum arguments on $\lambda,\,{\cal O},\,u_\nu$, as well as Dirac and color indices on $ \langle u_\nu \,{\cal O}\, \bar \lambda \rangle$, etc.
\item[3] The three point tree-level Green's function of the Gluino-Glue operator with an external ghost-antighost pair and a gluino vanishes.
\end{itemize}

Imposing renormalization conditions of the above two-point and three-point Green's functions is sufficient\footnote{One could of course calculate also four-point Green's functions; in doing so  a number of consistency checks would emerge regarding the divergent part of the mixing coefficients $z$. Further Green's functions (five-point and above) will bring in no superficial divergences.} in order to obtain the renormalization of the Gluino-Glue operator $Z_{Gg}$ and all mixing coefficients $z$. Once the renormalization factors in the $\MSbar$ scheme are determined, one can construct their RI$'$ counterparts using conversion factors which are immediately extracted from the above Green's function regularized in $DR$ to the required perturbative order. Being regularization independent, these same conversion factors can then be also used on the lattice. The same procedure can be applied in a straightforward manner to determine from our results the renormalization and mixing coefficients in other schemes, as well.

Our conventions for Fourier transformations are: 
\bea
u_\mu(x) &=& \int \frac{d^4q}{(2\pi)^4 } e^{i q \cdot x}\,\tilde{u}_\mu(q)\\%
\la(x) &=& \int \frac{d^4q}{(2\pi)^4 } e^{i q \cdot x}\,\tilde \la(q)\\
\bar\la(x) &=& \int \frac{d^4q}{(2\pi)^4 } e^{-i q \cdot x}\,\tilde{\bar{\la}}(q)\\
c(x) &=& \int \frac{d^4q}{(2\pi)^4 } e^{i q \cdot x}\,\tilde c(q) \\
\bar c(x) &=& \int \frac{d^4q}{(2\pi)^4 } e^{-i q \cdot x}\,\tilde{\bar{c}}(q)
\label{Fourier}
\eea
In what follows we will omit the tilde from Fourier-transformed fields. 

As already shown, in order to impose renormalization conditions, we need the expressions for certain tree-level Green's functions of the operators.  In particular, the nonvanishing two-point amputated tree-level Green's functions, with an operator insertion at point $x$, are:
\bea
\label{first2pt}
\langle u_\nu^{\alpha_1}(-q_1)\,{\cal O}_{Gg}\, \bar\lambda^{\alpha_2}(q_2) \rangle_{amp}^{tree}  &=& \frac{1}{2}\delta^{\alpha_{1}\alpha_{2}}ie^{i(q_{1}+q_{2})x}\sigma_{\mu \rho}(q_{1\mu}\delta_{\nu \rho}-q_{1\rho}\delta_{\mu \nu})\nonumber\\
&=&-\delta^{\alpha_{1}\alpha_{2}}ie^{i(q_{1}+q_{2})x}(\gamma_{\nu}\slashed q_{1}-q_{1\nu})\\
\langle u_\nu^{\alpha_1}(-q_1) \,{\cal O}_{A1}\,\bar\lambda^{\alpha_2}(q_2) \rangle _{amp}^{tree}  &=& \frac{1}{2}\delta^{\alpha_{1}\alpha_{2}}ie^{i(q_{1}+q_{2})x}q_{1\nu} \\
\langle u_\nu^{\alpha_1}(-q_1)\,{\cal O}_{B1}\,\bar\lambda^{\alpha_2}(q_2) \rangle _{amp}^{tree}  &=& \frac{1}{2}\delta^{\alpha_{1}\alpha_{2}}ie^{i(q_{1}+q_{2})x}(\gamma_{\nu}\gamma_{\rho})q_{2\rho}\\
\langle u_\nu^{\alpha_1}(-q_1)\,{\cal O}_{C1}\, \bar\lambda^{\alpha_2}(q_2) \rangle _{amp}^{tree}  &=& \frac{1}{2}\delta^{\alpha_{1}\alpha_{2}}ie^{i(q_{1}+q_{2})x}q_{2\nu} \\
\langle u_\nu^{\alpha_1}(-q_1) \,{\cal O}_{C2}\,\bar\lambda^{\alpha_2}(q_2) \rangle _{amp}^{tree}  &=& \frac{1}{2}\delta^{\alpha_{1}\alpha_{2}} e^{i(q_{1}+q_{2})x}\gamma_{\nu} 
\label{last2pt}
\eea
and the three-point amputated tree-level Green's functions of ${\cal O}_{Gg}$, ${\cal O}_{B1}$, ${\cal O}_{C3}$ and ${\cal O}_{C4}$:
\bea
\label{first3pt}
\langle u_\nu^{\alpha_1}(-q_1) \,u_\mu^{\alpha_2}(-q_2) \,{\cal O}_{Gg}\, \bar\lambda^{\alpha_3}(q_3) \rangle _{amp}^{tree}  &=& -g\,f^{\alpha_{1}\alpha_{2}\alpha_{3}}\,e^{i(q_{1}+q_{2}+q_{3})x}\sigma_{\nu \mu} = -g \,f^{\alpha_{1}\alpha_{2}\alpha_{3}}\,e^{i(q_{1}+q_{2}+q_{3})x}( \gamma_{\nu} \gamma_{\mu} - \delta_{\mu \nu} ) \\
\langle u_\nu^{\alpha_1}(-q_1) \,u_\mu^{\alpha_2}(-q_2) \,{\cal O}_{B1}\, \bar\lambda^{\alpha_3}(q_3) \rangle _{amp}^{tree}  &=& - g \,f^{\alpha_{1}\alpha_{2}\alpha_{3}}\,e^{i(q_{1}+q_{2}+q_{3})x}\sigma_{\nu \mu}  \\
\langle u_\nu^{\alpha_1}(-q_1) \,u_\mu^{\alpha_2}(-q_2) \,{\cal O}_{C3}\,\bar\lambda^{\alpha_3}(q_3) \rangle _{amp}^{tree}  &=& -g \,f^{\alpha_{1}\alpha_{2}\alpha_{3}}\,e^{i(q_{1}+q_{2}+q_{3})x}\sigma_{\nu \mu}  \\
\langle c^{\alpha_3}(q_3) \,{\cal O}_{A1}\,\bar c^{\alpha_2}(q_2) \,  \bar\lambda^{\alpha_1}(q_1)\rangle _{amp}^{tree}  &=& \frac{1}{2} g \,f^{\alpha_{1}\alpha_{2}\alpha_{3}}\,e^{i(q_{1}-q_{2}+q_{3})x}\\
\langle  c^{\alpha_3}(q_3) \,{\cal O}_{C4}\,\bar c^{\alpha_2}(q_2) \,  \bar\lambda^{\alpha_1}(q_1)\rangle _{amp}^{tree}  &=& - \frac{1}{2} g \,f^{\alpha_{1}\alpha_{2}\alpha_{3}}\,e^{i(q_{1}-q_{2}+q_{3})x}
\label{last3pt}
\eea

The structures on the rhs of Eqs.~(\ref{first2pt})-(\ref{last2pt}) are the only ones which may appear with divergent coefficients in the one-loop Green's function of ${\cal O}_{Gg}$: $\langle u_\nu^{\alpha_1}(-q_1)\,{\cal O}_{Gg}\, \bar\lambda^{\alpha_2}(q_2) \rangle_{amp}$; this allows us to determine unequivocally the coefficients $Z_{Gg}$, $z_{A1}$, $z_{B1}$, $z_{C1}$ and $z_{C2}$. The coefficients $z_{C3}$ and $z_{C4}$, which cannot be divergent, are determined by comparing the one-loop Green's functions: $\langle u_\nu^{\alpha_1}(-q_1) \,u_\mu^{\alpha_2}(-q_2) \,{\cal O}_{Gg}\, \bar\lambda^{\alpha_3}(q_3) \rangle _{amp}$ and $\langle  c^{\alpha_3}(q_3) \,{\cal O}_{Gg}\,\bar c^{\alpha_2}(q_2) \,  \bar\lambda^{\alpha_1}(q_1)\rangle _{amp}$ to the tree-level structures on the rhs of Eqs.~(\ref{first3pt})-(\ref{last3pt}).  

\section{Results at the continuum Regularization}
\label{sec4}
We use dimensional regularization in order to calculate the two-point and three-point Green's functions of ${\cal O}_{Gg}$ in the continuum, in $D = 4 - 2\epsilon$ dimensions.

There are in total 4 vertices in those diagrams of Fig.~\ref{fig2pt} which show up in the continuum. Two of them ($V^{\cal O}$) come from the operator ${\cal O}$ and the other two ($V^S$) from the action $S$. A factor $\int d^4 k/(2\pi)^4 \tilde X(k)$ is understood for each field $X$ appearing in the vertices. [In our conventions, indices for different fields appear in the following order: gluons, antigluinos, gluinos, antighosts, ghosts. Repeated indices are summed over. $k_j$ denote momenta; $\alpha_j, \beta$ are color indices in the adjoint representation; $\mu_j, \rho, \sigma$ are Lorentz indices. Furthermore, for compactness, vertices have not yet been symmetrized over identical particles.] 

The vertices of operator ${\cal O}_{Gg}$ with gluino/gluon fields and gluino/gluon/gluon fields are shown below. 
\bea
V^{{\cal O}; u, \la}_{\mu_1}(k_{1},k_{2})= \frac{i}{2}\delta^{\alpha_{1}\alpha_{2}}ie^{i(k_{1}+k_{2})x}\sigma_{\rho \sigma} (k_{1\rho}\delta_{\sigma \mu_1}-k_{1\sigma}\delta_{\rho \mu_1})
\eea
\bea
V^{{\cal O}; u, u, \la}_{\mu_1, \mu_2}(k_{1},k_{2},k_{3})=-\frac{g}{2}\,f^{\alpha_{1}\alpha_{2}\alpha_{3}}\,e^{i(k_{1}+k_{2}+k_{3})x}\sigma_{\rho \sigma}\delta_{\rho \mu_1}\delta_{\sigma \mu_2} 
\eea
Vertices coming from the continuum action, with gluino/antigluino/gluon fields and with three gluons are: \\
\be
V^{S; u, \bar \la, \la}_{\mu_1}(k_{1},k_{2},k_{3})=\frac{g}{2}(2\pi)^4 \delta(k_1-k_2+k_3)f^{\alpha_1\alpha_2\alpha_3} \gamma_{\mu_1}
\ee
\be
V^{S; u, u, u}_{\mu_1, \mu_2, \mu_3}(k_{1},k_{2},k_{3})=- i g (2\pi)^4\delta(k_1+k_2+k_3)f^{\alpha_1\alpha_2\alpha_3}\delta^{\mu_1\mu_2}
(- k_{1_{\mu_3}} + k_{2_{ \mu_3}})
\ee
Fig.~\ref{fig3ptguu} contains also the four-gluon action vertex:
\be
V^{S; u, u, u, u}_{\mu_1, \mu_2, \mu_3, \mu_4}(k_{1},k_{2},k_{3},k_{4}) = \frac{1}{4} g^2 (2\pi)^4 \delta(k_1+k_2+k_3+k_4) f^{\alpha_1\alpha_3\beta}f^{\beta \alpha_2 \alpha_4} \delta_{\mu_1 \mu_2}\delta_{\mu_3 \mu_4}
\ee
Finally, Fig.~\ref{fig3ptgCC} contains the ghost vertex:
\be
V^{S, u, \bar c, c}_{\mu_1}(k_{1},k_{2},k_{3}) = - i g (2\pi)^4 \delta(k_1-k_2+k_3) f^{\alpha_1\alpha_2\alpha_3} k_{2_{ \mu_1}}
\ee

To make use of Eq.~(\ref{2ptGFexpr}) we need to know also the factors $Z_u$ and $Z_\la$. For arbitrary values of $N_c$ and parameter $\beta$  ($N_f=0$) these are given by\footnote{We briefly recall the procedure for the extraction of these factors in appendix~\ref{appendA}.} :
\be
Z_u^{DR,\MSbar} = 1- \frac{g^2 N_c}{16\pi^2} \frac{1}{\epsilon} \left(1 + \frac{\beta}{2} \right)
\label{ZuDR}
\ee
\be
Z_\la^{DR,\MSbar} = 1+ \frac{g^2 N_c}{16\pi^2} \frac{1}{\epsilon} \left(1 -\beta  \right)
\label{ZlaDR}
\ee

The total expression for all Green's functions in $DR$ can be written as one part that contains the divergent terms (poles in $\epsilon$) and a second part with finite terms.  To return to four dimensions, we must be able to take the limit $\epsilon \to 0$. The $\MSbar$ renormalization scheme is set to eliminate the pole parts, leaving the finite terms intact. These terms make up the $\MSbar$ renormalized Green's functions and they will be used in order to extract the corresponding renormalization factors and mixing coefficients in the lattice regularization. In contrast, the $RI'$-like conditions eliminate the divergent part, but also alter the finite part.

Specifically, we calculate the two point Green's function of the Gluino-glue operator for the following three choices of momentum: $q_2=0$, $q_1=0$ and $q_2=-q_1$. For the choice $q_2=0$, we find:
\bea
\hspace{-.3cm} \langle u_\nu^{\alpha_1}(-q_1)  \,{\cal O}_{Gg}\, \bar\lambda^{\alpha_2}(q_2) \rangle _{amp}\big|^{DR}_{q_2 = 0}  &=&-\delta^{\alpha_{1}\alpha_{2}}ie^{iq_{1}x}(\gamma_{\nu}\slashed q_{1}-q_{1\nu}) + \frac{g^2 N_{c}}{16\pi^2}\frac{1}{2} \delta^{\alpha_1\,\alpha_2}e^{iq_1x}\nonumber \\&\times& \bigg[
i( \gamma_{\nu} \slashed q_1 - q_{1\nu}) \left(-\frac{12-3 \beta }{2 \epsilon} -6 -\beta + \frac{\beta^2}{2}  - \frac{12 -3 \beta }{2} \log\left(\frac{\bar \mu^2}{q_1^2}\right)\right) 
\bigg]
\label{GgCONTq2zero}
\eea
The pole part of this expression (actually also the finite part in this case) is proportional to the tree-level Green's function of ${\cal O}_{Gg}$ and thus there is no mixing with ${\cal O}_{A1}$: $z_{A1}^{DR,\MSbar} =0$. By imposing the renormalization condition of Eq.~(\ref{2ptGFexpr}) and demanding the lhs to be finite, $Z_{Gg}$ is determined to be:
\be
Z_{Gg}^{DR,\MSbar} = 1 - \frac{g^2 N_c}{16\pi^2} \frac{3}{\epsilon}
\ee
Indeed, $Z_{Gg}$ is gauge invariant in $\MSbar$. For the second choice of momentum ($q_1=0$), the tree-level Green's function of ${\cal O}_{Gg}$ gives zero, but the one-loop result is:
\bea
\langle u_\nu^{\alpha_1}(-q_1) \,{\cal O}_{Gg}\,\bar\lambda^{\alpha_2}(q_2) \rangle _{amp}\big|^{DR}_{q_1 = 0} &=&\frac{g^2 N_{c}}{16\pi^2}\frac{1}{2} \delta^{\alpha_1\,\alpha_2} e^{iq_2x}\bigg[-i q_{2 \nu} - i \gamma_{\nu} \slashed q_2 \left( \frac{3}{2 \epsilon} + 1 + \frac{3}{2} \log\left(\frac{\bar \mu^2}{q_2^2}\right) \right) \bigg]
\label{GgCONTq1zero}
\eea
The pole part of Eq.~(\ref{GgCONTq1zero}) determines immediately the mixing coefficient of ${{\cal O}_{B1}}$ in $DR$ and $\MSbar$:
\be
z_{B1}^{DR,\MSbar} =  \frac{g^2 N_c}{16\pi^2} \frac{3}{2\epsilon}
\ee
We note that this coefficient is also gauge independent, even though ${{\cal O}_{B1}}$ is a non-gauge invariant operator. The term proportional to the tree-level Green's function of the operator ${{\cal O}_{C1}}$ is finite and thus $z_{C1}^{DR,\MSbar}$ automatically vanishes. In the case of the lower dimension operator ${\cal O}_{C2}$, no mixing is expected to appear in the continuum, indeed $z_{C2}^{DR,\MSbar} = 0$. 

The last choice of momentum ($q_2 = -q_1$), for the two-point Green's function in $DR$, corresponds to the insertion of the Gluino-Glue operator at zero momentum. 
\bea
\langle u_\nu^{\alpha_1}(-q_1) \,{\cal O}_{Gg}\, \bar\lambda^{\alpha_2}(q_2) \rangle _{amp}\big|^{DR}_{q_2 = -q_1} &=& 
-\delta^{\alpha_{1}\alpha_{2}}i (\gamma_{\nu}\slashed q_{1}-q_{1\nu}) + \frac{g^2 N_{c}}{16\pi^2}\frac{1}{2} \delta^{\alpha_1\,\alpha_2}
\bigg[i(\gamma_{\nu} \slashed q_1 -q_{1\nu}) \bigg(- \frac{12 -3 \beta }{2 \epsilon} -7 + \frac{\beta^2}{2}  \nonumber \\
&&\hspace{1.5cm} -\frac{12 -3 \beta }{2}\log\bigg(\frac{\bar \mu^2}{q_1^2}\bigg ) \bigg )
+  i \gamma_{\nu} \slashed q_1 \bigg(\frac{3}{2 \epsilon} +2+ \frac{3}{2}\log\bigg(\frac{\bar \mu^2}{q_1^2}\bigg ) \bigg ) \bigg]
\label{GgCONTq1Pq2zero}
\eea
Eq.~(\ref{GgCONTq1Pq2zero}) is used as a consistency check: indeed its pole parts are eliminated upon applying the renormalization and mixing coefficients previously found.

Eliminating the pole parts of Eqs.~(\ref{GgCONTq2zero}), (\ref{GgCONTq1zero}) and (\ref{GgCONTq1Pq2zero}), one arrives at the $\MSbar$ renormalized two point Green's functions. The difference between the latter and the bare Green's functions on the lattice will give the corresponding renormalization factor and mixing coefficients on the lattice.

In order to determine the mixing of the remaining operators ${\cal O}_{C3}$, ${\cal O}_{C4}$ we have to calculate certain three-point Green's functions containing ${\cal O}_{Gg}$. Our result for the Green's function with external gluino, antighost and ghost fields is:
\be
\langle c^{a_3}(q_3) \,{\cal O}_{Gg}\,\bar c^{a_2}(q_2) \bar\lambda^{a_1}(q_1) \rangle _{amp}\big|^{DR}_{q_1=q_2,\, q_3 = 0}= \frac{g^2 N_{c}}{16\pi^2}\left(\frac{3}{4} (1 - \beta) g \,f^{\alpha_{1}\alpha_{2}\alpha_{3}}\,\right) 
\label{3ptGFgCCexp}
\ee
Eq.~(\ref{3ptGFgCCexp}) is necessarily pole free, since the tree-level value of this Green's function vanishes, and ${\cal O}_{C4}$ belongs to class C. Calculation of the same Green's function on the lattice will determine whether a (finite) mixing coefficient $z^{L,\MSbar}_{C4}$ will be necessary in order to match Eq.~(\ref{3ptGFgCCexp}).

From Eq.~(\ref{eptGFexprGGg}) we can verify that $z_{C3}$ also vanishes in $DR$. $Z_u^{DR,\MSbar}, Z_\la^{DR,\MSbar}, Z_{Gg}^{DR,\MSbar}$ and $Z_g^{DR,\MSbar}$ are required to eliminate the pole parts of the rhs of Eq.~(\ref{eptGFexprGGg}), leaving only finite parts. The lhs is actually the $\MSbar$-renormalized three point Green's function. The expression for $Z_g^{DR,\MSbar}$ is (see, e.g., Ref.~\cite{Costa:2017rht} for $N_f=0$):
\be
Z_{g}^{DR,\MSbar}=1 + \frac{g^2\,}{16\,\pi^2} \frac{1}{\epsilon} \frac{3}{2} N_c.
\ee

In contrast to Eq.~(\ref{3ptGFgCCexp}) which is finite, the bare three-point Green's function with an external gluino and two gluons is not. The contributions from the diagrams of Fig.~\ref{fig3ptguu}, taken separately, are not proportional to tree-level. However, their sum has this property and it takes the following form in the continuum:

\begin{eqnarray}
\langle u_\nu^{\alpha_1}(-q_1) u_\mu^{\alpha_2}(-q_2) \,{\cal O}_{Gg}\, \bar\lambda^{\alpha_3}(q_3) \rangle _{amp}\big|^{DR}_{q_2 =0 , q_3 = -q_1} &=& - g \,f^{\alpha_{1}\alpha_{2}\alpha_{3}}\,( \gamma_{\nu} \gamma_{\mu} - \delta_{\mu \nu} ) Z_{g}^{-1}\nonumber\\ 
&&\hspace{-5cm}+ \frac{g^3 N_c}{16\pi^2} \,f^{\alpha_{1}\alpha_{2}\alpha_{3}}\, \Bigg[\delta_{\mu \nu} \left(\frac{81}{16} +\frac{5}{2 \epsilon} -\frac{\beta}{\epsilon} -\frac{1}{4} \beta^2 -\frac{5}{8} \beta +\frac{5}{2} \log \left(\frac{\bar \mu^2}{q_1^2}\right) - \beta \log \left(\frac{\bar \mu^2}{q_1^2}\right)\right) \nonumber\\
&&\hspace{-5cm}+ \gamma_\nu \gamma_\mu \left(-3-\frac{5}{2\epsilon}+\frac{\beta}{\epsilon} +\frac{\beta}{4} +\frac{\beta^2}{4}+\beta \log \left(\frac{\bar \mu^2}{q_1^2}\right)-\frac{5}{2} \log \left(\frac{\bar \mu^2}{q_1^2}\right)\right) \nonumber\\
&&\hspace{-5cm}+  \gamma_\nu \frac{\slashed q_1 q_{1 \mu}}{q_1^2} \left(\frac{77}{16} -\frac{13 \beta }{8} +\frac{\beta^2}{4}\right)+ \gamma_\mu \frac{\slashed q_1 q_{1 \nu}}{q_1^2}   \left(\frac{1}{16} -\frac{\beta }{4} \right) -\frac{ q_{1 \nu} q_{1 \mu}}{q_1^2} \left(\frac{63}{8} - \frac{9\beta}{4} +\frac{\beta ^2}{4}\right)
\Bigg]
\end{eqnarray}

In the above equation the terms proportional to $1/\epsilon$ cancel against the renormalization factors of the fields, of the Gluino-Glue operator, of the coupling constant and of the mixing with the operator ${\cal O}_{B1}$. Therefore, using the condition of Eq.~(\ref{eptGFexprGGg}), mixing with ${\cal O}_{C3}$ is not observed and thus $z_{C3}^{DR,\MSbar}=0$.
\vspace*{0.5cm}

\section{Lattice Regularization}
\label{sec5}
In our lattice calculation, we extend Wilson's formulation of the QCD action, to encompass SUSY partner fields as well. In this standard discretization, gluinos reside on the lattice sites, and gluons reside on the links of the lattice: $U_\mu (x) \equiv U_{x,x+\mu}= e^{i g a T^{\alpha} u_\mu^\alpha (x+a\hat{\mu}/2)}$ where $\alpha$ is a color index in the adjoint representation of the gauge group and $a$ is the lattice spacing. This formulation leaves no SUSY generators intact \cite{Bergner:2016sbv}, and it also breaks chiral symmetry; it thus represents a ``worst case'' scenario, which is worth investigating in order to address the complications \cite{Giedt:2009yd} which will arise in numerical simulations of SUSY theories. In our ongoing investigation we plan to address also improved actions \cite{Neuberger:1997fp, Luscher:1998}, so that we can check to what extent some of the SUSY breaking effects can be alleviated. The gluinos are described by clover improved Wilson fermions in the adjoint representation and the Euclidean action ${\cal S}^{L}_{\rm SYM}$ on the lattice becomes\cite{Ali:2018dnd}:
\begin{equation}
\begin{split}
{\cal S}^{L}_{\rm SYM}&=a^{4}\sum_{x}\bigg[\frac{N_{c}}{g^{2}}\sum_{\mu, \nu}\bigg(1-\frac{1}{N_{c}}TrU_{\mu \nu}\bigg)+ \sum_{\mu}\bigg(Tr\bigg(\bar\lambda\gamma_{\mu} D_{\mu} \lambda\bigg)-\frac{a r}{2}Tr\bigg(\bar\lambda D^{2}\lambda \bigg)\bigg)-\sum_{\mu, \nu}\bigg(\frac{c_{\rm SW} \ a}{4}\bar\lambda^{\alpha}\sigma_{\mu \nu}\hat{\tilde{F}}_{\mu \nu}^{\alpha \beta}\lambda^{\beta}\bigg)\bigg]\\
\end{split}
\label{susylagrLattice}
\end{equation}
where $\hat{\tilde{F}}_{\mu \nu}^{ab}$ in the adjoint representation is defined as:

\bea
\hat{\tilde{F}}_{\mu \nu}^{\alpha \beta}&=&\frac{1}{8}(\tilde{Q}_{\mu \nu}^{\alpha \beta}-\tilde{Q}_{\nu \mu}^{\alpha \beta})\\
\tilde{Q}_{\mu \nu}^{\alpha \beta}&=&2{\rm{tr}}_c \bigg( 
T^\alpha\, U_{x,x+\mu}U_{x+\mu,x+\mu+\nu}U_{x+\mu+\nu,x+\nu}U_{x+\nu,x}T^\beta\,U_{x,x+\nu}U_{x+\nu,x+\mu+\nu}U_{x+\mu+\nu,x+\mu}U_{x+\mu,x} \nonumber\\
&&\phantom{{\rm{tr}}_c } + T^\alpha\, U_{x,x+\nu}U_{x+\nu,x+\nu-\mu}U_{x+\nu-\mu,x-\mu}U_{x-\mu,x} T^\beta\, U_{x,x+\mu}U_{x+\mu,x+\mu-\nu}U_{x+\mu-\nu,x-\mu}U_{x-\nu,x} \nonumber\\
&&\phantom{{\rm{tr}}_c } + T^\alpha\, U_{x,x-\mu}U_{x-\mu,x-\mu-\nu}U_{x-\mu-\nu,x-\nu}U_{x-\nu,x} T^\beta\, U_{x,x-\nu}U_{x-\nu,x-\mu-\nu}U_{x-\mu-\nu,x-\mu}U_{x-\mu,x}\nonumber\\
&&\phantom{{\rm{tr}}_c} + T^\alpha\, U_{x,x-\nu}U_{x-\nu,x-\nu+\mu}U_{x-\nu+\mu,x+\mu}U_{x+\mu,x} T^\beta\, U_{x,x-\mu}U_{x-\mu,x-\mu+\nu}U_{x-\mu+\nu,x+\nu}U_{x+\nu,x}\bigg)
\eea
and
\begin{equation}
U_{\mu \nu}(x) =U_\mu(x)U_\nu(x+a\hat\mu)U^\dagger_\mu(x+a\hat\nu)U_\nu^\dagger(x)
\end{equation}
The 4-vector $x$ is restricted to the values $x = na$, with $n$ being an integer 4-vector. The terms proportional to the Wilson parameter, $r$, eliminate the problem of fermion doubling, at the expense of breaking chiral invariance\footnote{In what follows, we will set $|r|=1$.}. In the limit $a \to 0$ the classical lattice action reproduces the continuum one. A gauge-fixing term, together with the compensating ghost field term, must also be added to the action, in order to avoid divergences from the  integration over gauge orbits; these terms are the same as in the non-supersymmetric case. Similarly, a standard ``measure'' term must be added to the action, in order to account for the Jacobian in the change of integration variables: $U_\mu \to u_\mu$\,. All the details and definitions of the continuum and the lattice actions can be found in Ref.\cite{Costa:2017rht}.

The definitions of the covariant derivatives are as follows:
\bea
{\cal{D}}_\mu\lambda(x) &\equiv& \frac{1}{2a} \Big[ U_\mu (x) \lambda (x +  a \hat{\mu}) U_\mu^\dagger (x) - U_\mu^\dagger (x - a \hat{\mu}) \lambda(x - a \hat{\mu}) U_\mu(x - a \hat{\mu}) \Big] \\
{\cal D}^2 \lambda(x) &\equiv& \frac{1}{a^2} \sum_\mu \Big[ U_\mu (x)  \lambda (x + a \hat{\mu}) U_\mu^\dagger (x)  - 2 \lambda(x) +  U_\mu^\dagger (x - a \hat{\mu}) \lambda (x - a \hat{\mu}) U_\mu(x - a \hat{\mu})\Big]
\eea

The Gluino-Glue operator on the lattice is defined as:
\be
{\cal O}_{Gg} = \sigma_{\mu \nu} {\rm{tr}}_c (\, \hat{F}_{\mu \nu}\lambda )
\label{GgO}
\ee
where
\begin{equation}
\begin{split}
\hat{F}_{\mu \nu}&=\frac{1}{8}(Q_{\mu \nu}-Q_{\nu \mu})\\
Q_{\mu \nu}&=U_{x,x+\mu}U_{x+\mu,x+\mu+\nu}U_{x+\mu+\nu,x+\nu}U_{x+\nu,x}\\
\quad& +U_{x,x+\nu}U_{x+\nu,x+\nu-\mu}U_{x+\nu-\mu,x-\mu}U_{x-\mu,x}\\
\quad& +U_{x,x-\mu}U_{x-\mu,x-\mu-\nu}U_{x-\mu-\nu,x-\nu}U_{x-\nu,x}\\
\quad& +U_{x,x-\nu}U_{x-\nu,x-\nu+\mu}U_{x-\nu+\mu,x+\mu}U_{x+\mu,x}
\end{split}
\end{equation}


Lattice vertices are very lengthy and are not presented here for the sake of brevity\footnote{Vertices are available from the authors upon request.}. Some of the vertices have no analog in the continuum; although these vertices vanish in the continuum limit, they contribute beyond tree level in perturbation theory even in the limit $a \to 0$.

For completeness, we present all relevant two- and three-point Green's functions, shown in Eqs.~(\ref{GgLATTq2zero}),~(\ref{GgLATTq1zero}),~(\ref{GgLATTq1Pq2zero}) and (\ref{GgLATT3ptGGgq3zero}), on the lattice. The renormalization conditions which we impose involve the renormalization factors of the gluino ($Z_{\lambda}$), gluon ($Z_{u}$), ghost ($Z_{c}$) fields and of the coupling constant ($Z_{g}$). Since we used the clover action for gluino fields, $Z_{\lambda}$ and $Z_{u}$ are recalculated, leading to\footnote{For brevity, decimal numbers in our results are presented only with four digits after the decimal point; they are known to higher accuracy.}:
\begin{eqnarray}
Z_{\lambda}^{L,\overline{\textrm{MS}}} &=&  1 - \frac{g^2\,N_c}{16\,\pi^2} \bigg(12.8524 + 3.7920 \beta - 5.5891\, c_{\rm SW}^2 - 4.4977\, c_{\rm SW} r + (1-\beta) \log(a^2\,\bar\mu^2) \bigg)
\label{ZlaL}
\end{eqnarray}
\begin{eqnarray}
Z_{u}^{L,\overline{\textrm{MS}}} &=&  1 + \frac{g^2\,N_c}{16\,\pi^2} \bigg[19.7392\frac{1}{N_c^2}- 17.1775 - 1.3863  \beta + 18.8508\, c_{\rm SW}^2-1.5939\, c_{\rm SW}r + \left(1+\frac{\beta}{2}\right) \log(a^2\,\bar\mu^2)\bigg)
\label{ZuL}
\end{eqnarray}
which coincide with the expressions in Ref.~\cite{Costa:2017rht} for $c_{\rm SW}=0$ and $N_f=0$. Divergences in renormalization factors manifest themselves as logarithms in the lattice spacing. The calculation of $Z_{\lambda}$ and $Z_{u}$ as well as the critical value for the gluino mass are presented in appendix~\ref{appendA}. Further, in Ref.~\cite{Costa:2017rht}, the ghost and the coupling constant renormalizations were presented for Wilson fermions and gluons. $Z_{c}^{L,\MSbar}$ is the same here because the ghost propagator does not involve gluino fields, and therefore the clover parameter does not appear in its expression. On the other hand $Z_g^{L,\MSbar}$, since it is calculated from the gluon-ghost-antighost Green's function, is changed here due to the presence of the clover term in $Z_{u}^{L,\MSbar}$. The new value of $Z_g^{L,\MSbar}$ is:
\begin{eqnarray}
Z_{g}^{L,\overline{\textrm{MS}}} &=& 1 + \frac{g^2\,}{16\,\pi^2}\Bigg[ -9.8696 \frac{1}{N_c}+ N_c\Big(12.8904 + 0.7969 \, c_{\rm SW}\ r - 9.4254 \, c_{\rm SW}^2 - \frac{3}{2}\log(a^2\,\bar\mu^2)\Big)\Bigg]
\end{eqnarray}

The computation of the bare Green's functions of ${\cal O}_{Gg}$ on the lattice is the most demanding part of the present work. The algebraic expressions involved are split into two parts: a) A part that can be evaluated in the $a \to 0$ limit: It contains terms which have a complicated dependence on the external momentum $q$ and show up in the regularization independent renormalized Green's functions. b) Terms which are divergent as $a \to 0$; their dependence on $q$ is necessarily polynomial. Our computations were performed in a covariant gauge, with arbitrary value of the gauge parameter $\beta$. Both renormalized and bare lattice Green's functions have the same tensorial forms, but the bare ones have additional lattice contributions. 

The first two-point Green's function for $q_2=0$ (cf. Eqs.~(\ref{2ptGFexpr}), (\ref{GgCONTq2zero})) will provide us with the renormalization of the Gluino-Glue operator, since it is proportional to its tree-level value:
\begin{eqnarray}
\langle u^{\alpha_{1}} (-q_1)\,{\cal O}_{Gg}\,\bar\lambda^{\alpha_{2}}_\nu(q_2) \rangle_{amp}\big|^L_{q_2 = 0} &=&-\delta^{\alpha_{1}\alpha_{2}}ie^{iq_{1}x}(\gamma_{\nu}\slashed q_{1}-q_{1\nu}) \nonumber\\ 
&+&\frac{g^2N_c}{16\pi^2}\frac{1}{2}\delta^{\alpha_1\, \alpha_2}\,e^{iq_1x}\,
i\,(\gamma_{\nu}\slashed q_1 - q_{1\nu})\bigg(\frac{-39.4784}{N_c^2}+27.5552+4.1783\beta\nonumber\\ 
&+&\frac{1}{2}\beta^2- 4.6002 \ {c_{\rm SW}}^2 
-12.8568 \ c_{\rm SW} \ r +6\log(a^2 q_1^2)-\frac{3\beta}{2}\log(a^2 q_1^2)\bigg)\nonumber\\
\label{GgLATTq2zero}
\end{eqnarray}
The determination of the renormalization factor, $Z_{Gg}^{L, \MSbar}$, follows by imposing the renormalization condition of Eq.~(\ref{2ptGFexpr}), in which the lhs is the $\MSbar$ renormalized Green's function (Eq.~(\ref{GgCONTq2zero}) without the pole terms). Our result is:
\begin{eqnarray}
Z_{Gg}^{L, \MSbar} = 1 - \frac{g^2 N_c}{16\pi^2}\bigg( \frac{9.8696}{N_c^2}-1.7626-9.9198\, c_{\rm SW}^2+4.9765\, c_{\rm SW}\, r-3\log(a^2\,\bar\mu^2)\bigg)
\end{eqnarray}

The same Green's function, evaluated at $q_1=0$, provides the mixing coefficient with the operator ${\cal O}_{B1}$ in accordance with Eq.~(\ref{2ptGFexpr}).
\begin{eqnarray}
\langle u^{a_{1}} (-q_1)\,{\cal O}_{Gg}\,\bar\lambda^{a_{2}}_\nu(q_2) \rangle_{amp}\big|^{L}_{q_1 = 0} &=&\frac{g^2 N_{c}}{16\pi^2}\frac{1}{2} \delta^{\alpha_1\,\alpha_2}\,e^{iq_2x}\, \bigg[-i q_{2 \nu} - i \gamma_{\nu} \slashed q_2 \left( 1.42407 - \frac{3}{2} \log(a^2 q_2^2) \right) \bigg]\nonumber\\
\label{GgLATTq1zero}
\end{eqnarray}
By comparing the finite parts of Eq.~(\ref{GgCONTq1zero}) with the lattice Green's function Eq.~(\ref{GgLATTq1zero}), the coefficient of $\gamma_{\nu} \slashed q_2$ determines the mixing coefficient with ${\cal O}_{B1}$. 
\begin{eqnarray}
z_{B1}^{L, \overline{\textrm{MS}}} =  \frac{g^2 N_c}{16\pi^2}\bigg(0.4241 - \frac{3}{2}\log(a^2\,\bar\mu^2)\bigg)
\end{eqnarray}

An immediate check of our results is the extraction of the $\MSbar$-renormalized Green's function at $q_2=-q_1$, followed by a comparison with our continuum result, shown in Eq.~(\ref{GgCONTq1Pq2zero}). This can be easily done by applying $Z_{Gg}^{L, \MSbar}$ and $z_{B1}^{L, \overline{\textrm{MS}}}$ in the condition of Eq.~(\ref{2ptGFexpr}), and using the bare lattice Green's function at $q_2=-q_1$: 
\begin{eqnarray}
\langle u_\nu^{\alpha_{1}}(-q_1) \,{\cal O}_{Gg}\, \bar\lambda^{\alpha_{2}}(q_2) \rangle _{amp}\big|^L_{q_2 = -q_1} &=&\frac{g^2N_c}{16\pi^2}\frac{1}{2}\delta^{\alpha_1\, \alpha_2}\bigg[
i(\gamma_{\nu}\slashed q_1 - q_{1\nu})\bigg(\frac{-39.4784}{N_c^2}+26.5552+5.1783\beta\nonumber\\ 
&+&\frac{1}{2}\beta^2- 4.6002\,{c_{\rm SW}}^2 
-12.8568\,c_{\rm SW} \ r+6\log(a^2 q_1^2)-\frac{3\beta}{2} \log(a^2 q_1^2)\bigg) \nonumber\\ 
&+&i\gamma_{\nu}\slashed{q_1}\bigg(2.4241-\frac{3}{2} \log(a^2 q_1^2)\bigg)\bigg] 
\label{GgLATTq1Pq2zero}
\end{eqnarray}

For all other operators, having non vanishing tree-level Green's functions with one external gluon and one external gluino field, their mixing coefficients automatically vanish: $z_{A1}^{L,\MSbar} = z_{C1}^{L,\MSbar} = z_{C2}^{L,\MSbar}=0$. 

We now turn to the three-point Green's functions. As we have already mentioned, the lattice three-point Green's function, $\langle c^{\alpha_3}(q_3) \,{\cal O}_{Gg}\,\bar c^{\alpha_2}(q_2) \bar\lambda^{\alpha_1}(q_1) \rangle _{amp}\big|^{L}_{q_1=q_2,\, q_3 = 0}$ coincides with the one in the continuum:
\be
 \langle c^{\alpha_3}(q_3) \,{\cal O}_{Gg}\,\bar c^{\alpha_2}(q_2) \bar\lambda^{\alpha_1}(q_1) \rangle _{amp}\big|^{L}_{q_1=q_2,\, q_3 = 0} =  \langle c^{\alpha_3}(q_3) \,{\cal O}_{Gg}\,\bar c^{\alpha_2}(q_2) \bar\lambda^{\alpha_1}(q_1) \rangle _{amp}\big|^{DR}_{q_1=q_2,\, q_3 = 0} = \frac{g^2 N_{c}}{16\pi^2}\left(\frac{3}{4} (1 - \beta) g \,f^{\alpha_{1}\alpha_{2}\alpha_{3}}\,\right)
\ee
It follows that $z_{C4}^{L,\MSbar}$ vanishes. On the other hand for the lattice Green's function with one external gluino and two external gluons, we find:
\begin{eqnarray}
\langle u_\nu^{\alpha_1}(-q_1) u_\mu^{\alpha_2}(-q_2) \,{\cal O}_{Gg}\, \bar\lambda^{\alpha_3}(q_3) \rangle _{amp}\big|^L_{q_2 =0 , q_3 = -q_1} &=& - g \,f^{\alpha_{1}\alpha_{2}\alpha_{3}}\,( \gamma_{\nu} \gamma_{\mu} - \delta_{\mu \nu} ) Z_{g}^{-1} \nonumber\\
&&\hspace{-5cm} - 39.4784 \frac{g^3}{16\pi^2 N_c} \,f^{\alpha_{1}\alpha_{2}\alpha_{3}}\,( \gamma_{\nu} \gamma_{\mu} - \delta_{\mu \nu} )\nonumber\\
&&\hspace{-5cm} +\frac{g^3 N_c}{16\pi^2} \,f^{\alpha_{1}\alpha_{2}\alpha_{3}}\, \Bigg[ \delta_{\mu \nu} \Big(-28.2259 - 0.08447 \beta - \frac{1}{4} \beta^2 + 6.4284\, c_{\rm SW}\, r + 2.3001 \, c_{\rm SW}^2  \nonumber\\
&&\hspace{-2cm} - \frac{5}{2} \log \left( a^2 q_1^2 \right) + \beta \log \left( a^2 q_1^2 \right)\Big) \nonumber\\
&&\hspace{-5cm} +\gamma_\nu \gamma_\mu \Big(30.2884 - 0.2905 \beta +\frac{\beta^2}{4} - 6.4284\, c_{\rm SW} r - 2.3001 \, c_{\rm SW}^2 \nonumber\\   
&&\hspace{-2cm} - \beta \log \left(a^2 q_1^2 \right) + \frac{5}{2} \log \left(a^2 q_1^2\right) \Big) \nonumber\\
&& \hspace{-5cm} \gamma_\nu \frac{\slashed q_1 q_{1 \mu}}{q_1^2} \left(\frac{77}{16} -\frac{13 \beta }{8} +\frac{\beta^2}{4}\right)+ \gamma_\mu \frac{\slashed q_1 q_{1 \nu}}{q_1^2}   \left(\frac{1}{16} -\frac{\beta }{4} \right)-\frac{ q_{1 \nu} q_{1 \mu}}{q_1^2} \left(\frac{63}{8} - \frac{9\beta}{4} +\frac{\beta ^2}{4}\right)
\Bigg]
\label{GgLATT3ptGGgq3zero}
\end{eqnarray}
The difference between the $\MSbar$ renormalized and bare Green's functions consists only of expressions proportional to the tree-level Green's functions of operators ${\cal O}_{Gg}$,  ${\cal O}_{B1}$ and  ${\cal O}_{C3}$; in this way, the rhs of Eq.~(\ref{eptGFexprGGg}) can be rendered equal to the corresponding lhs, by an appropriate definition of the renormalization factors and mixing coefficients on the lattice. Indeed, taking this difference removes the following structures: $q_\nu q_\mu/q^2$, $\gamma_\nu \slashed q q_\mu /q^2$ and $\gamma_\mu \slashed q q_\nu/q^2$ from Eq.~(\ref{GgLATT3ptGGgq3zero}) leaving only contributions proportional to the tensorial structure: $( \gamma_{\mu} \gamma_{\nu} - \delta_{\mu \nu})$. By using the one-loop renormalization factors of the fields, the coupling constant renormalization and operators ${\cal O}_{Gg}$,  ${\cal O}_{B1}$ and  ${\cal O}_{C3}$, we end up with a linear equation whose only unknown is the mixing coefficient, $z_{C3}^{L,\MSbar}$. Note that $Z_{Gg}^{L,\MSbar}$ and $Z_{B1}^{L,\MSbar}$, which we find using the two-point Green's functions, along with the renormalization of fields and coupling constant, do render Eq.~(\ref{GgLATT3ptGGgq3zero}) finite as expected. We find for $z_{C3}^{L,\MSbar}$:
\be
z_{C3}^{L,\MSbar} =  \frac{g^2}{16\pi^2}\bigg( \frac{-19.7392}{N_c} + N_c \left(20.3883 - 3.8228 \beta  \right) \bigg)
\ee
The above result is independent of the choice of the clover parameter, $c_{\rm SW}$.

\section{Summary and Future Plans}
\label{sec6}
In this paper, we performed a detailed perturbative study of the Gluino-Glue operator. This operator is directly connected to light bound states of the theory, and its renormalization is very important as a necessary step forward for extracting the spectrum of low-lying bound states from numerical simulations.

Our study of the Gluino-Glue operator entails a two-step regularization procedure: 

(i) continuum regularization, where we calculate Green's functions of the Gluino-Glue operator in order to derive the $\MSbar$ renormalized Green's functions. We provide all the bare Green's functions in $DR$; from these the reader may straightforwardly determine the conversion factors to other schemes, such as $RI'$ which can be used nonperturbatively. The calculation of conversion factors from $\MSbar$ to a new Gauge Invariant Renormalization Scheme (GIRS) in coordinate space is presently underway. Within GIRS one obtains the renormalization factor of ${\cal O}_{Gg}$ nonperturbatively. 

(ii) the calculation of lattice regularized Green's functions: this is the most demanding part of the present work. We determine perturbatively the renormalization factors and all mixing coefficients. Use of these quantities converts bare Green's functions, calculated in lattice simulations, directly to $\MSbar$, without need for an intermediate scheme such as $RI'$.

It will be interesting to study properties of the fermionic Gluino-Glue particle, in parallel with bosonic glueballs and mesonic gluinoballs. The Gluino-Glue particle and gluinoballs are expected to be the SUSY partners of glueballs, and therefore it is important to verify via simulations that these particles have the same mass (if we recover a non-broken phase of SUSY in the continuum limit). A similar investigation in this direction is to study a three-gluino operator, $~f^{\alpha_{1}\alpha_{2}\alpha_{3}}\,  \lambda^{\alpha_1} (\bar \lambda^{\alpha_2} \Gamma \lambda^{\alpha_3})$ in order to explore baryonic states in SYM.

In the future, we aim to compute the renormalization of the supercurrent~${\rm{tr}}_c(\bar \lambda \gamma^{\mu} \sigma^{\rho \nu} u_{\rho \nu})$ in the context of SYM theory. The renormalization of the supercurrent and its potential mixing also require the calculation of Green's functions with an external gluino and gluons. A further direction in this context regards supersymmetric Ward identities in order to study the lattice artifacts and the recovery of supersymmetry in the continuum limit~\cite{Ali:2020mvj}.
Finally, we plan to carry out extensive perturbative study of the Gluino-Glue operator and of the supercurrent in Supersymmetric QCD (SQCD). The Green's functions of the above operators in SQCD exhibits a very complicated mixing pattern under renormalization, involving also squark and quark fields. 

\appendix
\section{Perturbative one-loop Renormalization of $Z_u$ and $Z_\lambda$ on the Lattice in the $\MSbar$ scheme}
\label{appendA}

The renormalization for the gluon field, $Z_u$, in the continuum, can be evaluated from the gluon propagator $\langle    u^{\alpha_1}_\mu(q_1)   u^{\alpha_2}_\nu(q_2) \rangle^{DR}_{\rm{inv}}$.

The corresponding one-loop Feynman diagrams are shown in Fig.~\ref{gluon2pt}. Their contributions, taken separately, are not transverse; however their sum does have this property. We find:

\begin{figure}[ht!]
\centering
\includegraphics{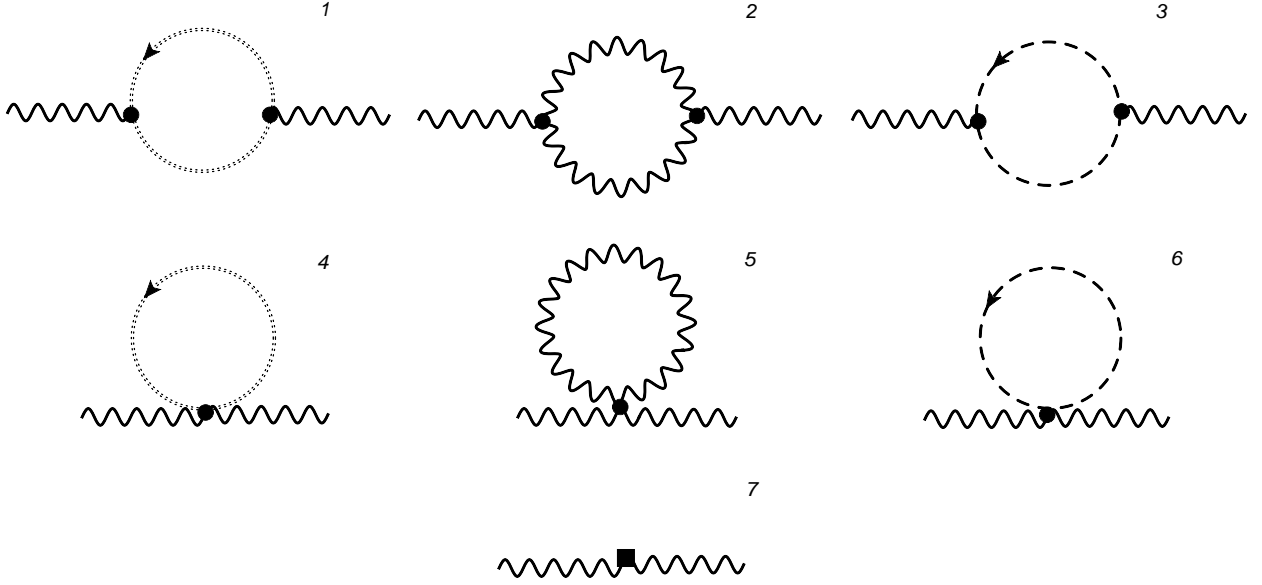}
\caption{One-loop Feynman diagrams contributing to the two point Green's function  $\langle  u_\mu^{\alpha_1}(q_1) u_\nu^{\alpha_2}(q_2) \rangle$.
  A wavy (dashed) line represents gluons (gluinos).  The ``double dashed'' line is the ghost field. Only the first three diagrams appear in $DR$.}
\label{gluon2pt}
\end{figure}

\bea
\langle    u^{\alpha_1}_\mu(q_1)   u^{\alpha_2}_\nu(q_2) \rangle^{DR}_{\rm{inv}} &=&(2\pi)^4 \delta(q_1+q_2) \delta^{\alpha_1\, \alpha_2} \Bigg\{  \frac{1}{1-\beta} q_{1\mu} q_{1\nu}\nonumber\\
&&\hspace{-0.5cm} + \left(q_1^2 \delta_{\mu \nu} - q_{1\mu} q_{1\nu}\right)\Bigg[ 1 - \frac{g^2\, N_c}{16\,\pi^2}\frac{1}{2} \left(\left(2+\beta\right)\frac{1}{\epsilon} + \frac{14}{3} - 2 \beta + \frac{\beta^2}{2} + \left(2+\beta\right)\log\left(\frac{\bar\mu^2}{q_1^2} \right)\right)\Bigg] \Bigg\}\label{gluonPropDR}
\eea
The one-loop result for $Z_u^{DR,\overline{\textrm{MS}}}$ (Eq.~(\ref{ZuDR})) follows directly from the above.

Since there is no one-loop longitudinal part for the gluon self-energy, the renormalization factor for the gauge parameter receives no one-loop contribution.\\ 
 On the lattice, all seven diagrams appearing in Fig.~\ref{gluon2pt} contribute to the gluon one-loop inverse propagator. We find:
\bea
\langle    u^{a_1}_\mu(q_1)   u^{a_2}_\nu(q_2) \rangle^{L}_{\rm{inv}} &=&(2\pi)^4 \delta(q_1+q_2) \delta^{a_1\, a_2} \Bigg\{ \frac{1}{1-\beta}  q_{1\mu} q_{1\nu}\\\nonumber
&&+ \left(q_1^2 \delta_{\mu \nu} - q_{1\mu} q_{1\nu}\right)\Bigg[1- \frac{g^2}{16\,\pi^2}\Big[ -19.7392 \frac{1}{N_c} \\\nonumber
&&+ N_c\left(19.5109 + 0.386294\beta +  \frac{\beta^2}{4} +  1.59389 \,c_{\rm SW} r- 18.8508 \,c_{\rm SW}^2 - \left(1+\frac{\beta}{2}\right)  \log\left(a^2\, q_1^2 \right)\right)\Big]\Bigg]\Bigg\}.
\label{GF2gluonlatt}
\eea
We notice that this result is proportional to the tree-level two-point Green's function of the gluon field. Some diagrams contribute a quadratically divergent mass term ($1/a^2$ contribution). But when all Feynman diagrams are summed these divergences are found to cancel out, as expected from gauge invariance. By demanding the following:
\be
\langle u_\mu^R\,u_\nu^R \rangle _{\rm inv} = Z_{u}^{-1}\, \langle u_\mu^B\,u_\nu^B \rangle^L_{\rm inv},
\label{GG2gluoncondition}
\ee
we find $Z_{u}^{L,\overline{\textrm{MS}}}$ (Eq.~({\ref{ZuL})).

We turn now to $Z_{\lambda}$. The one-loop diagrams contributing to the inverse gluino propagator are shown in Fig.~\ref{gluino2pt}.

\begin{figure}[ht!]
\centering
\includegraphics[scale =0.75]{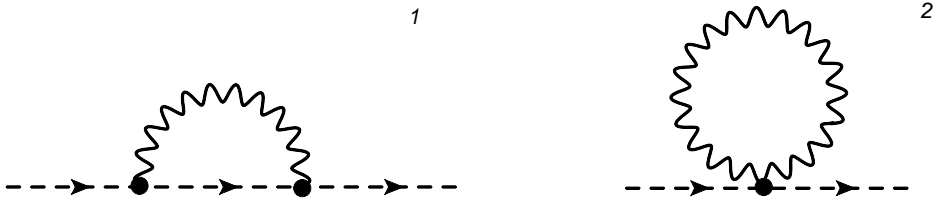}
\caption{One-loop Feynman diagrams contributing to the two point Green's function  $\langle  \lambda^{\alpha_1}(q_1) \bar \lambda^{\alpha_2}(q_2) \rangle$. A wavy (dashed) line represents gluons (gluinos). Only the first diagram appears in $DR$.}
\label{gluino2pt}
\end{figure}

In $DR$, we find:
\bea
\langle   \lambda^{\alpha_1}(q_1)  {\bar{\lambda}}^{\alpha_2}(q_2) \rangle^{DR}_{\rm{inv}} &=& (2\pi)^4 \delta(q_1-q_2) \frac{i}{2}\,\delta^{\alpha_1\, \alpha_2} \slashed{q_1} \Bigg[ 1 + \frac{g^2\, N_c}{16\,\pi^2} \left(1-\beta + \frac{(1-\beta)}{\epsilon} + (1-\beta) \log\left(\frac{\bar\mu^2}{q_1^2}  \right)\right)\Bigg].
\eea
The $DR$ renormalization factor of the gluino field in the $\MSbar$ scheme follows directly (see Eq.~(\ref{ZlaDR})).

On the lattice the gluino inverse propagator is given at the one-loop order by:
\bea
\label{GF2gluinolatt}
\langle   \lambda^{\alpha_1}(q_1)  {\bar{\lambda}}^{\alpha_2}(q_2) \rangle^{L}_{\rm{inv}} &=& (2\pi)^4 \delta(q_1-q_2)\delta^{\alpha_1\, \alpha_2} \Bigg\{ \frac{i}{2} \, \slashed{q_1} \Bigg[ 1 - \frac{g^2\, N_c}{16\,\pi^2} \Big[11.8524 +  4.7920 \beta - 4.4977 \, c_{\rm SW} r - 5.5891 \, c_{\rm SW}^2  \nonumber\\
&&\hspace{5cm}+ (1-\beta) \log\left(a^2\, q_1^2 \right)\Big]\Bigg] \nonumber\\
&&\hspace{3cm} + \frac{g^2}{16\,\pi^2} \frac{N_c}{2} \frac{1}{a} \left( 51.4347 \,r  - 27.4663 \, c_{\rm SW} -22.8606 \, c_{\rm SW}^2 r \right)\Bigg\}.
\eea

The renormalization factor of the gluino field is determined in the $\MSbar$ scheme by imposing the condition:
\be
\langle\lambda^R\,{\bar{\lambda}}^{R} \rangle _{\rm inv} = Z_{\lambda}^{-1}\, \langle\lambda^B\,{\bar{\lambda}}^{B} \rangle^L_{\rm inv}
\label{GG2gluinocondition}
\ee
leading to $Z_{\lambda}^{L,\overline{\textrm{MS}}}$ (Eq.~(\ref{ZlaL})).
The terms in the last line of Eq.~(\ref{GF2gluinolatt}) have a power divergence in $a$. They dictate the ``critical mass'' which must be included in the bare Lagrangian in order to obtain a massless renormalized gluino propagator. Hence, the critical mass for the gluino field is:
\be
m^{gluino}_{crit.} =  \frac{g^2\,N_c }{16\,\pi^2}\ \,\frac{1}{a}\left(  51.4347 \,r - 27.4663 \, c_{\rm SW} -22.8606\, c_{\rm SW}^2 \,r \right)
\ee

\begin{acknowledgements}
M.C. and H.P. acknowledge financial support from the project ``Quantum Fields on the Lattice'', funded by the Cyprus Research and Innovation Foundation (RIF) under contract number EXCELLENCE/0918/0066.
\end{acknowledgements}


\begin{thebibliography}{99}

\bibitem{Martin:1997ns}
S.~P.~Martin,
``A Supersymmetry primer'',
Adv. Ser. Direct. High Energy Phys. \textbf{21} (2010), 1-153
[arXiv:hep-ph/9709356 [hep-ph]].

\bibitem{Quevedo:2010ui}
  F.~Quevedo, S.~Krippendorf and O.~Schlotterer,
``Cambridge Lectures on Supersymmetry and Extra Dimensions'',
  arXiv:1011.1491 [hep-th], and references therein.

\bibitem{Zyla:2020}
P.~A.~Zyla et al. (Particle Data Group), Prog. Theor. Exp. Phys. 2020, 083C01 (2020).

\bibitem{Santra:2020mfi}
  A.~Santra [ATLAS Collaboration],
 ``QCD Issues in Searches for Supersymmetry with the ATLAS Detector'',
  Nucl.\ Part.\ Phys.\ Proc.\  {\bf 309-311} (2020) 49.

\bibitem{CMS:2019tlp}
A.~M.~Sirunyan \textit{et al.} [CMS Collaboration],
``Search for supersymmetry in pp collisions at $\sqrt{s}=$ 13 TeV with 137 fb$^{-1}$ in final states with a single lepton using the sum of masses of large-radius jets'',
Phys. Rev. D \textbf{101} (2020) no.5, 052010
[arXiv:1911.07558 [hep-ex]].

\bibitem{Curci:1986sm} 
G.~Curci, G.~Veneziano,
``Supersymmetry and the Lattice: A Reconciliation?'',
Nucl. Phys. B {\bf 292} (1987) 555.

\bibitem{Creutz:2001}
M.~Creutz, 
``Aspects of chiral symmetry and the lattice'', 
Rev. Mod. Phys. {\bf 73} (2001) 119 [arXiv:hep-lat/0007032]. 


\bibitem{Giet&Poppitz}
J.~Giedt, E.~Poppitz, 
``Lattice supersymmetry, superfields and renormalization'', 
JHEP {\bf 9} (2004) 029 [arXiv:hep-th/0407135]. 

\bibitem{Kaplan:2009}
S.~Catterall, D.B.~Kaplan, M.~\"Unsal,
``Exact lattice supersymmetry'',
Phys. Rep. {\bf 484} (2009) 71 [arXiv:0903.4881 [hep-lat]].

\bibitem{Catterall:2014vga}
S.~Catterall, J.~Giedt, D.~Schaich, P.~H.~Damgaard and T.~DeGrand,
``Results from lattice simulations of N=4 supersymmetric Yang--Mills'',
PoS LATTICE {\bf 2014} (2014) 267
[arXiv:1411.0166 [hep-lat]].

\bibitem{Joseph:2015xwa}
A.~Joseph, 
``Review of Lattice Supersymmetry and Gauge-Gravity Duality'',
Int.\ J.\ Mod.\ Phys.\ A {\bf 30} (2015) 1530054
[arXiv:1509.01440 [hep-th]].

\bibitem{Ali:2018fbq}
S.~Ali, H.~Gerber, I.~Montvay, G.~M\"unster, S.~Piemonte, P.~Scior, G.~Bergner,
``Analysis of Ward identities in supersymmetric Yang-Mills theory'',
Eur.\ Phys.\ J.\ C {\bf 78} (2018) 404
[arXiv:1802.07067 [hep-lat]].

\bibitem{Endrodi:2018ikq}
  G.~Endrodi, M.~Kaminski, A.~Schafer, J.~Wu and L.~Yaffe, ``Universal Magnetoresponse in QCD and $\mathcal{N}=4$ SYM'',
  JHEP {\bf 1809} (2018) 070
  [arXiv:1806.09632 [hep-th]].


\bibitem{Giedt:2009yd}
J.~Giedt,
``Progress in four-dimensional lattice supersymmetry'',
Int. J. Mod. Phys. A \textbf{24} (2009), 4045
[arXiv:0903.2443 [hep-lat]].

\bibitem{Bergner:2016sbv}
G.~Bergner and S.~Catterall,
``Supersymmetry on the lattice'',
Int. J. Mod. Phys. A \textbf{31} (2016) no.22, 1643005
[arXiv:1603.04478 [hep-lat]].

\bibitem{Ali:2020mvj}
  S.~Ali, G.~Bergner, H.~Gerber, I.~Montvay, G.~M\"unster, S.~Piemonte and P.~Scior,
  ``Continuum extrapolation of Ward identities in ${{\mathcal {N}}=1}$ supersymmetric SU(3) Yang-Mills theory'',
  Eur.\ Phys.\ J.\ C {\bf 80} (2020) no.6,  548
  [arXiv:2003.04110 [hep-lat]].


\bibitem{Costa:2017rht}
M.~Costa and H.~Panagopoulos,
``Supersymmetric QCD on the Lattice: An Exploratory Study'',
Phys. Rev. D \textbf{96} (2017) no.3, 034507
[arXiv:1706.05222 [hep-lat]].

\bibitem{Costa:2018mvb}
M.~Costa and H.~Panagopoulos,
``Supersymmetric QCD: Renormalization and Mixing of Composite Operators'',
Phys. Rev. D \textbf{99} (2019) no.7, 074512
[arXiv:1812.06770 [hep-lat]].

\bibitem{Ali:2018dnd}
S.~Ali, G.~Bergner, H.~Gerber, P.~Giudice, I.~Montvay, G.~M\"unster, S.~Piemonte and P.~Scior,
``The light bound states of $\mathcal{N}=1$ supersymmetric SU(3) Yang-Mills theory on the lattice'',
JHEP \textbf{03} (2018), 113
[arXiv:1801.08062 [hep-lat]].

\bibitem{Ali:2019agk}
S.~Ali, G.~Bergner, H.~Gerber, I.~Montvay, G.~M\"unster, S.~Piemonte and P.~Scior,
``Numerical results for the lightest bound states in $\mathcal{N}=1$ supersymmetric SU(3) Yang-Mills theory'',
Phys. Rev. Lett. \textbf{122} (2019) no.22, 221601
[arXiv:1902.11127 [hep-lat]].

\bibitem{VEN} G. Veneziano and S. Yankielowicz, ``An Effective Lagrangian for the Pure N=1 Supersymmetric Yang-Mills Theory'', Phys. Lett. B 113 (1982) 231.

\bibitem{Collins:1984xc}
J.~C.~Collins,
``Renormalization'',
Cambridge University Press (2010).

\bibitem{Collins:1994ee}
J.~C.~Collins and R.~J.~Scalise,
``The Renormalization of composite operators in Yang-Mills theories using general covariant gauge'',
Phys. Rev. D \textbf{50} (1994), 4117
[arXiv:hep-ph/9403231 [hep-ph]].

\bibitem{Miller:1983pg}
R.~D.~C.~Miller,
``Supersymmetric gauge fixing and the effective potential'',
Phys. Lett. B \textbf{129} (1983), 72.

\bibitem{Neuberger:1997fp}
H.~Neuberger,
``Exactly massless quarks on the lattice'',
Phys. Lett. B \textbf{417} (1998), 141
[arXiv:hep-lat/9707022].


\bibitem{Luscher:1998}
M.~L\"uscher,
``Exact chiral symmetry on the lattice and the Ginsparg-Wilson relation'',
 Phys. Lett. B {\bf 428} (1998) 342 [arXiv:hep-lat/9802011].


\end{thebibliography}
\end{document}